\documentclass[12pt]{article}
\usepackage{epsfig,amsmath,amssymb}
\usepackage{graphicx,psfrag} 
\usepackage{mathtools}
\hypersetup{linktocpage}
\numberwithin{equation}{section}

\newcommand{\be}{\begin{equation}}
\newcommand{\ee}{\end{equation}}
\newcommand{\bea}{\begin{eqnarray}}
\newcommand{\eea}{\end{eqnarray}}
\newcommand{\bA}{\begin{array}}
\newcommand{\eA}{\end{array}}
\newcommand{\bc}{\begin{center}}
\newcommand{\ec}{\end{center}}
\newcommand{\al}{\alpha}

\newcommand{\ra}{\rightarrow}
\newcommand{\del}{\partial}

\newcommand{\ie}{{\it i.e.}}
\newcommand{\eg}{{\it e.g.}}

\newcommand{\lan}{\langle}
\newcommand{\ran}{\rangle}
\newcommand{\ua}{\uparrow}
\newcommand{\da}{\downarrow}

\begin{document}


\begin{titlepage}
\vspace{25mm}

\bc

\hfill 
\\         [20mm]

{\Huge $N$-level ghost-spins and entanglement}  \vspace{1mm}
\vspace{10mm}

{\large  Dileep P. Jatkar$^{1,2}$, Kedar S. Kolekar$^3$, and K.~Narayan$^3$} \\
\vspace{3mm}
{\small \it 1. Harish-Chandra Research Institute\\}
{\small \it Chhatnag Road, Jhusi, Allahabad 211019, India.\\[3mm]}

{\small \it 2. Homi Bhabha National Institute\\}
{\small \it Training School Complex, Anushakti Nagar, Mumbai 400085,
  India.\\[3mm]}

{\small \it 3. Chennai Mathematical Institute, \\ }
{\small \it SIPCOT IT Park, Siruseri 603103, India.}

\ec
\medskip
\vspace{40mm}

\begin{abstract}
  Ghost-spins, 2-level spin-like variables with indefinite norm have
  been studied in previous work. Here we explore various $N$-level
  generalizations of ghost-spins. First we discuss a flavoured
  generalization comprising $N$ copies of the ghost-spin system, as
  well as certain ghost-spin chains which in the continuum limit lead
  to 2-dim $bc$-ghost CFTs with $O(N)$ flavour symmetry. Then we
  explore a symplectic generalization that involves antisymmetric
  inner products, and finally a ghost-spin system exhibiting $N$
  irreducible levels.  We also study entanglement properties. In all
  these cases, we show the existence of positive norm ``correlated
  ghost-spin'' states in two copies of ghost-spin ensembles obtained
  by entangling identical ghost-spins from each copy: these exhibit
  positive entanglement entropy.

\end{abstract}

\end{titlepage}


\newpage 
{\footnotesize
\begin{tableofcontents}
\end{tableofcontents}}


\section{Introduction}

Ghost-spin systems \cite{Narayan:2016xwq,Jatkar:2016lzq,Jatkar:2017jwz}
and their patterns of entanglement are interesting from multiple
points of view.  Some of these possible applications involve ghost
sectors in theories with gauge symmetry, while others pertain to
conjectures involving de Sitter physics and $dS/CFT$ dualities
\cite{Strominger:2001pn,Witten:2001kn,Maldacena:2002vr,Anninos:2011ui,
  Bousso:2001mw,Balasubramanian:2002zh,Harlow:2011ke,Ng:2012xp,Das:2012dt,
  Anninos:2012ft,Anninos:2017eib} including higher spin
versions. Ghost-spins are 2-level spin-like variables but with
indefinite norm \cite{Narayan:2016xwq}: thus in some ways, they are
best regarded as simple quantum mechanical toy models for theories
with negative norm states. Ghost-spin systems have interesting
entanglement patterns by virtue of this indefinite norm.  If one
considers ghost spin systems with even number of ghost spins then it
is possible to find subspaces in the Hilbert space where one obtains
positive entanglement entropy for positive norm states
\cite{Narayan:2016xwq,Jatkar:2016lzq}.  The situation is not so
fortuitous in the case of odd number of ghost spins.  However, several
interesting physical systems which contain indefinite norm states seem
to admit even numbers of such states.  Taking the study of the ghost
spins further, it was shown in \cite{Jatkar:2017jwz} that appropriate
ghost-spin chains in the continuum limit give rise to the $bc$-ghost
CFTs in two dimensions. This suggests that they may be regarded as
microscopic building blocks for ghost-CFTs and perhaps more general
non-unitary theories.

The explorations of the ghost-spin system so far have used 2-state
spin-like variables with indefinite inner product.  It is interesting
to study generalizations of ghost-spins with flavour degrees of
freedom assigned to them.  These flavour quantum numbers may be
relevant for applications to non-abelian gauge theories. However, the
ghost spin system with flavours by themselves is an interesting set up
worthy of exploration.  Once we allow for flavour symmetry, there are
multiple ways in which they can be incorporated in the ghost-spins
framework.  One way is to assign an index, say $A$, which takes $N$
values, to a 2-component ghost-spin with either $O(N)$ or $Sp(N)$
symmetry in the flavour index.  The inner product in the flavour space
is given by $\delta^{AB}$ for the $O(N)$ case and $\Omega^{AB}$ in the $Sp(N)$ case. More general inner products can also be analyzed.

In this paper we will explore these generalizations of the 2-state
ghost-spin system to $N$-level systems.  In
sec.~\ref{sec:ghost-spins-n}, we will begin with a brief recap of the
known results followed by a list of the $N$-level generalizations of
ghost-spins that we discuss in this paper.  We introduce the $O(N)$
and $Sp(N)$ inner products in the flavour indices as well as more
general inner products $J^{AB}$ which are symmetric but non-vanishing
for $A\neq B$.  In sec.~\ref{sec:n-level-ghost}, we look at the
$O(N)$ generalization of the 2-level system where we denote the
flavoured ghost spins in terms of the 2-component spins carrying an
additional index corresponding to the global $O(N)$ symmetry,
$|\uparrow^A\rangle$.  We write the indefinite inner product between
spins by splitting it into indefinite product between
$|\uparrow\rangle$ and $|\downarrow\rangle$ and a symmetric product
$\delta_{AB}$ in the $O(N)$ index $A$. This as we will see is
analogous to the $bc$-ghost system studied in the flavourless 2-level
system.  We then consider the ghost-spin chain with $O(N)$ flavour
symmetry and show that it leads to the flavoured $bc$-ghost CFTs
(sec.~\ref{sec:gschainO(N)}).  We then consider correlated ghost-spin
states in sec.~\ref{sec:CorrStO(N)} and analyse the entanglement
pattern in them.  We find that the results that we had obtained in the
even ghost-spin system in the flavourless case carry over to the case
with flavours.  We generically find a subspace of correlated
ghost-spin states in the Hilbert space with positive norm states
having positive entanglement.  Finally we briefly comment on more
general inner products which lead to spin-glass type couplings in
sec.~\ref{sec:symmetric-spin-glass}. We then turn our attention in
sec.~\ref{sec:sympl} to symplectic inner products between certain
generalizations of ghost-spins and study entanglement in correlated
ghost spin states.  Finally in sec.~\ref{sec:n-irreducible-levels} we
consider a generalization of 2-level ghost-spins to $N$ irreducible
levels: this is slightly different from the flavoured generalizations
above. We study the entanglement pattern in correlated ghost-spin
states here as well.  In sec.~\ref{sec:discussion}, we summarise our
results and comment on their applications to the dS/CFT
correspondence, in part reviewing the picture in
\cite{Narayan:2017xca} of $dS_4$ as approximately dual to a
thermofield-double type entangled state between two copies of
ghost-CFTs.

\section{Ghost-spins and $N$-level generalizations}\label{sec:ghost-spins-n}

Before getting to $N$-level generalizations, we first briefly review
some essential aspects of ghost-spins.  Ghost-spins were defined in
\cite{Narayan:2016xwq} as simple toy quantum mechanical models for
theories with negative norm states, abstracting from $bc$-ghost
CFTs. These constructions were motivated by the studies
\cite{Narayan:2015vda,Narayan:2015oka} on certain complex extremal
surfaces in de Sitter space with negative area in $dS_4$ which amount
to analytic continuations of the Ryu-Takayanagi formulations of
holographic entanglement entropy
\cite{Ryu:2006bv,Ryu:2006ef,HRT,Rangamani:2016dms}.
In contrast with a single spin which has\
$\lan\ua|\ua\ran = 1 = \lan\da|\da\ran$,\ 
a single ghost-spin is defined as a 2-state spin variable with
indefinite inner product
\be
\lan\ua|\da\ran = 1 = \lan\da|\ua\ran\ , \qquad
\lan\ua|\ua\ran = 0 = \lan\da|\da\ran\ .
\ee
A general state $\psi^+|+\ran+\psi^-|-\ran$ thus has norm
$|\psi^+|^2-|\psi^-|^2$, which is not positive definite.
By changing basis, the states
$|\pm\ran={1\over\sqrt{2}}(|\ua\ran\pm |\da\ran)$ have manifestly
positive/negative norm, satisfying $\lan\pm |\pm\ran=\pm 1$.
We can then normalize general positive/negative norm states with norm
$\pm 1$ respectively.
Consider now a state comprising two ghost-spins: this has norm
\be\label{norm2gs}
|\psi\ran=\psi^{\al\beta}|\al\beta\ran:\qquad
\lan\psi|\psi\ran=\gamma_{\alpha\kappa} \gamma_{\beta\lambda}
\psi^{\alpha\beta} {\psi^{\kappa\lambda}}^*\ ,\qquad \gamma_{++}=1,\ \ 
\gamma_{--}=\lan -|-\ran=-1\ ,
\ee
where $\gamma_{\al\beta}$ is the indefinite metric. Thus although states
$|-\ran$ have negative norm, the state $|-\ran|-\ran$ has positive
norm. The full density matrix is
$\rho=|\psi\rangle\langle\psi| = \sum \psi^{\alpha\beta} {\psi^{\kappa\lambda}}^* 
|\alpha\beta \rangle\langle \kappa\lambda|$.
Tracing over one of the ghost-spins leads to a reduced density matrix
$(\rho_A)^{\alpha\kappa} = \gamma_{\beta\lambda} \psi^{\alpha\beta} {\psi^{\kappa\lambda}}^* = \gamma_{\beta\beta}  \psi^{\alpha\beta} {\psi^{\kappa\beta}}^*$,
\bea\label{rhoA2gs}
\qquad\qquad 
(\rho_A)^{++} =\ |\psi^{++}|^2 - |\psi^{+-}|^2\ , &\quad&
(\rho_A)^{+-} =\ \psi^{++} {\psi^{-+}}^* - \psi^{+-} {\psi^{--}}^*\ , 
\nonumber\\
(\rho_A)^{-+} =\ \psi^{-+} {\psi^{++}}^* - \psi^{--} {\psi^{+-}}^*\ , &\quad& 
(\rho_A)^{--} =\ |\psi^{-+}|^2 - |\psi^{--}|^2\ , 
\eea
for the remaining ghost-spin.
Then $tr \rho_A = \gamma_{\alpha\kappa} (\rho_A)^{\alpha\kappa} =
(\rho_A)^{++} - (\rho_A)^{--}$. 
Thus the reduced density matrix is normalized to have\ 
$tr \rho_A = tr \rho = \pm 1$ depending on whether the state 
(\ref{norm2gs}) is positive or negative norm.
The entanglement entropy calculated as the von Neumann entropy of 
$\rho_A$ is\ $S_A = -\gamma_{\alpha\beta} (\rho_A \log \rho_A)^{\alpha\beta}$,
perhaps best defined using a mixed-index reduced density matrix
$(\rho_A)^\alpha{_\kappa} = \gamma_{\beta\kappa} (\rho_A)^{\al\beta}$.
This can be illustrated via a simple family of states
\cite{Narayan:2016xwq} with a diagonal reduced density matrix: setting
${\psi^{-+}}^* = \psi^{+-} {\psi^{--}}^*/\psi^{++}$\ in the states
(\ref{norm2gs}) gives\
\bea\label{Ex:2gsEE}
&& (\rho_A)^{\al\beta}|\al\ran\lan\beta| = \pm x |+ \rangle\langle +|\
\mp\ (1-x) |- \rangle\langle -|\ , \qquad
x = {|\psi^{++}|^2\over |\psi^{++}|^2 + |\psi^{--}|^2} \qquad [0 < x < 1] ,
\nonumber\\
&& (\rho_A)_\alpha^\kappa = \gamma_{\alpha\beta} (\rho_A)^{\beta\kappa}:
\qquad (\rho_A)^+_+ = \pm x ,\qquad (\rho_A)^-_- = \pm (1-x) ,\quad
\eea
where the $\pm$ pertain to positive/negative norm states respectively\
(note that\ $tr\rho_A = (\rho_A)^+_+ + (\rho_A)^-_- = \pm 1$).
The location of the negative eigenvalue is different for
positive/negative norm states, leading to different results for the
von Neumann entropy. Now $\log\rho_A$ simplifies to\  
$(\log\rho_A)^+_+ = \log (\pm x)$ and $(\log\rho_A)^-_- = \log (\pm (1-x))$.
The entanglement entropy defined as\
$S_A = -\gamma_{\alpha\beta}(\rho_A\log\rho_A)^{\alpha\beta}$ becomes\
$S_A = - (\rho_A)^+_+ (\log\rho_A)^+_+ - (\rho_A)^-_- (\log\rho_A)^-_-$\ 
and so
\be\label{Ex:2gsEE2}
\langle\psi| \psi\rangle \gtrless 0:\qquad
S_A = - (\pm x)\log (\pm x) - (\pm (1-x)) \log (\pm (1-x)) \ .
\ee
For positive norm states, $S_A$ is manifestly positive since $x<1$,
just as in an ordinary 2-spin system.\ Negative norm states give a 
negative real part for EE since $x<1$ and the logarithms are negative:
further there is an imaginary part (the simplest branch has
$\log (-1)=i\pi$).\

Now consider restricting to the subspace\
\be\label{2gsCorrSt}
|\psi\ran=\psi^{++}|++\ran+\psi^{--}|--\ran\quad \longrightarrow\quad
\lan\psi|\psi\ran = |\psi^{++}|^2+|\psi^{--}|^2 > 0\ .
\ee
These states of ``correlated ghost-spins'' comprise entanglement
between two copies of identical states: they can be seen to be strictly
positive norm, with a positive reduced density matrix (\ref{rhoA2gs})
and positive entanglement. In \cite{Narayan:2017xca}, a picture of de
Sitter space as a thermofield double type state (with de Sitter
entropy then emerging as the entanglement entropy) was discussed based
on such correlated ghost-spin states in two copies of ghost-CFTs at
the future and past boundaries of $dS_4$ in the static coordinatization.
For ensembles with an even number of ghost-spins, such correlated
ghost-spin states always exist comprising positive norm subsectors, as
argued in \cite{Jatkar:2016lzq}, where ensembles of ghost-spins were
developed further with regard to their entanglement properties. Odd
ghost-spins were found to behave differently: for instance,\
$|\psi\ran=\psi^{++\ldots}|++\ldots\ran+\psi^{--\ldots}|--\ldots\ran$
has norm
$\lan\psi|\psi\ran=|\psi^{++\ldots}|^2 + (-1)^n|\psi^{--\ldots}|^2$\
and mixed-index RDM components\
$(\rho_A)^+_+=|\psi^{++\ldots}|^2,\ (\rho_A)^-_-=(-1)^n|\psi^{--\ldots}|^2$.
This is not positive definite for $n$ odd (even if
$\lan\psi|\psi\ran>0$).  Ensembles of ghost-spins and spins were also
found to exhibit interesting entanglement patterns.

In \cite{Jatkar:2017jwz}, certain 1-dim ghost-spin chains with specific
nearest-neighbour interactions were found to yield $bc$-ghost CFTs in
the continuum limit, \ie\ these ghost-spin chains are in the same
universality class as those ghost-CFTs. We will not review this here
since this will effectively be encompassed in a related detailed
description later.

\vspace{1mm}

\noindent\underline{\bf $N$-level ghost-spins:}\ \ In this paper, we
generalize the $2$-level ghost-spin reviewed above to $N$-levels by
considering various generalizations as outlined below:\\
\noindent $\bullet$\ \ $O(N)$ symmetry flavour generalization of the
$bc$-ghost system:
\be\label{NdeltaAB}
\langle\downarrow^A|\uparrow^B\rangle=\delta^{AB}=\langle\uparrow^A|\downarrow^B\rangle\ , \qquad \langle\downarrow^A|\downarrow^B\rangle=\langle\uparrow^A|\uparrow^B\rangle=0\ , \qquad A,B=1,2,\dots,N\ .
\ee
These are essentially $N$ copies of the 2-level ghost-spin system.
It is then possible to find appropriate ghost-spin chains which lead
to a generalization of the $bc$-ghost system but with internal $O(N)$
flavour indices.\ \ A simple generalization of this case involves the
flavours having a spin-glass type interaction with coupling $J_{AB}$\
which is in general non-vanishing for $A\neq B$,
\be\label{NJAB}
\langle\downarrow^A|\uparrow^B\rangle=J^{AB}=\langle\uparrow^A|\downarrow^B\rangle\ ; \qquad
\langle\downarrow^A|\downarrow^B\rangle=\langle\uparrow^A|\uparrow^B\rangle=0\ , \qquad A,B=1,2,\dots,N\ .
\ee
In flavour space, this thus encodes possibly nonlocal flavour couplings.
Taking the $J_{AB}$ matrix to be real and symmetric allows diagonalization
and in that diagonal basis, this can be reduced to the
above $O(N)$ flavoured case.\\
$\bullet$\ \ $N$-levels with symplectic-like structure:
\be\label{NOmegaAB}
\langle\uparrow^A|\downarrow^B\rangle=i\,\Omega^{AB}\ , \quad \langle\downarrow^A|\uparrow^B\rangle=i\,\Omega^{AB}\ , \quad \langle\uparrow^A|\uparrow^B\rangle=0=\langle\downarrow^A|\downarrow^B\rangle\ , \quad\ A,B=1,\dots,2N\ .
\ee
These have a symplectic structure built into the inner product, which
was in part motivated by 3-dim ghost-CFTs of symplectic fermions
\cite{LeClair:2006kb,LeClair:2007iy} that have been discussed in the
conjectured duals to higher spin $dS_4$ \cite{Anninos:2011ui}.\\
$\bullet$\ \ $N$ irreducible levels, \ie\ we generalize the two states
$|\uparrow\rangle, |\downarrow\rangle$ to $|e_1\rangle,\dots,|e_N\rangle$
such that
\be\label{Nirred}
\langle e_i|e_i\rangle=0\ ;  \quad \langle e_i|e_j\rangle=1 \quad \text{for}\ i\neq j, \quad i,j=1,2,\dots,N\ .
\ee
This case is slightly different from the previous cases in that the
elemental ghost-spins are not 2-level anymore (with flavour indices),
but irreducibly $N$-level.

In all these cases representing $N$-level generalizations of
ghost-spin ensembles, we will argue that correlated ghost-spin states
exist comprising a uniformly positive norm subspace of states with the
interpretation of entanglement between two copies of the state space.
This will be the main point of the paper.

It is possible to find operator realizations consistent with some of
these inner products above. The first case essentially comprises $N$
copies of the $bc$-operator algebra,
\be
\{ \sigma_b^A, \sigma_c^B\} = \delta^{AB}\ .
\ee
We can then define ghost-spin-chain Hamiltonians with nearest
neighbour hopping type interactions but with the flavours decoupled,
\be
H = \sum_n \left( \sigma_{bn}^A\sigma_{c(n+1)}^B
+ \sigma_{b(n+1)}^A\sigma_{cn}^B \right) \delta_{AB}
\quad\xrightarrow{\ \ ?\ \ }\quad \int b^A\del c^A\ .
\ee
Based
on the fact that the continuum limit for the single flavour case is the
familiar $bc$-ghost CFT \cite{Jatkar:2017jwz}, the continuum limit can
be argued to be flavoured generalizations of $bc$-ghost CFTs, with the
flavour contractions exhibiting $O(N)$ symmetry.
We will discuss this in detail in sec.~3.
The symplectic inner products above are consistent with the operator
algebra
\be
\{ \sigma_b^A, \sigma_c^B\} = i\Omega_{AB}\ ,
\ee
as we will discuss in sec.~5. The continuum limit is less clear in
this case, although there are indications that these may be related
to logarithmic CFTs \cite{Gurarie:1993xq,Kausch:1995py,Gurarie:1997dw,
  Kausch:2000fu,Flohr:2001zs,Krohn:2002gh,Gurarie:2013tma}.




\section{$N$-level ghost-spins with $O(N)$ flavour symmetry}\label{sec:n-level-ghost}

In this section, we consider an $N$-level generalization of
ghost-spins with $O(N)$ symmetry among the $N$ internal flavour
indices, defined by (\ref{NdeltaAB}), \ie\ we have the elemental
inner products
\bea
&& \langle\uparrow^A|\downarrow^B\rangle = \delta^{AB} 
= \langle\downarrow^A|\uparrow^B\rangle , \quad
\langle\uparrow^A|\uparrow^B\rangle=0=\langle\downarrow^A|\downarrow^B\rangle ,
\qquad A,B=1,2,\dots,N\ ,\nonumber\\
&& \qquad
|\pm^A\rangle=\frac{1}{\sqrt{2}}(|\uparrow^A\rangle\pm|\downarrow^A\rangle)\ ,
\qquad \langle \pm^A|\pm^B\rangle=\pm\delta^{AB}\ , \qquad
\langle \pm^A|\mp^B\rangle=0\ . \label{O(N)elementalproducts}
\eea
This is essentially $N$ copies of the 2-level ghost-spin reviewed in
sec.~2. In the second line, we have defined a convenient basis where
the inner product is diagonal: this makes manifest the negative norm
basis states.

Consider first a single ghost-spin with $N$ flavours.
The general configuration is defined by specifying the simultaneuous
configurations for each of the $N$ flavours so the general state
comprising various basis states $|s_i\ran$ is
\be\label{basisStsiON}
|\psi\ran = \psi^{s_i}|s_i\ran\ ,\qquad\quad
|s_i\ran \equiv \left\{\ \left|\! {\tiny \bA{c}
  \ua^1\\ \ua^2\\ \ua^3\\  \vdots \eA} \!\right\ran,\ \
\left|\! {\tiny \bA{c}
  \da^1\\ \ua^2\\ \ua^3\\  \vdots \eA} \!\right\ran,\ \
\left|\! {\tiny \bA{c}
  \ua^1\\ \da^2\\ \ua^3\\  \vdots \eA} \!\right\ran,\ \
\ldots\ \ \right\}\ ,
\ee
\ie\ in the first basis state, the first flavour is $\ua^1$, the
second is $\ua^2$, third being $\ua^3$ and so on, and likewise for the
other basis states. It is important to note that the $|s_i\ran$ are
really direct product states over the various flavour components:
although we have written them as column vectors for convenience of
notation (especially in light of the discussion later on multiple
ghost-spins), the inner products between these states is not a dot
product between two column vectors.
Instead we define the inner products between the configurations $|s_i\ran$
as
\begin{equation}\label{flavouredinnerproduct}
\lan s_i|s_j\ran = \frac{1}{N!}\sum \epsilon_{A_1 A_2 \dots A_N}\epsilon_{B_1 B_2 \dots B_N}\lan s^{A_1}_i|s^{B_1}_j\ran \lan s^{A_2}_i|s^{B_2}_j\ran \cdots \lan s^{A_N}_i|s^{B_N}_j\ran\ ,
\end{equation}
where $i,j=1,2,\dots,2^N$ label the configurations, $A_1, B_1, \dots
=1,2,\dots,N$ label the flavours and $\epsilon_{A_1 A_2 \dots A_N}$ is
the totally symmetric tensor with $\epsilon_{12\dots N}=1$ and
$\epsilon_{A_1 A_2 \dots A_N}$ vanishes if any two labels are the same.
In other words, $\epsilon_{A_1 A_2 \dots A_N}$ is simply a book-keeping
device for ensuring that each elemental state $|s_j^A\ran$ in $|s_j\ran$
is paired with another corresponding elemental state in $\lan s_i|$.

To illustrate how this works, let us consider a simple example of a
single ghost-spin with $N=2$ flavours: the distinct configurations of
this system are described by the basis states
\be\label{statesO(N)-siN=2}
|s_1\ran = \big|^{\ua^1}_{\ua^2}{\big\ran}\ ,\qquad
|s_2\ran = \big|^{\ua^1}_{\da^2}{\big\ran}\ ,\qquad
|s_3\ran = \big|^{\da^1}_{\ua^2}{\big\ran}\ ,\qquad
|s_4\ran = \big|^{\da^1}_{\da^2}{\big\ran}\ .
\ee
Then the inner product \eqref{flavouredinnerproduct} simplifies to
\begin{equation}\label{flavouredinnerproductN=2}
\lan s_i|s_j\ran = \frac{1}{2!}\sum \epsilon_{A_1 A_2}\epsilon_{B_1 B_2}\lan s^{A_1}_i|s^{B_1}_j\ran \lan s^{A_2}_i|s^{B_2}_j\ran = \lan s^{1}_i|s^{1}_j\ran \lan s^{2}_i|s^{2}_j\ran + \lan s^{1}_i|s^{2}_j\ran \lan s^{2}_i|s^{1}_j\ran\ ,
\end{equation}
where we have used $\epsilon_{12}=1=\epsilon_{21}$. 
Using the elemental inner products in \eqref{O(N)elementalproducts} gives
\begin{equation}
\lan s_i|s_j\ran = \lan s^{1}_i|s^{1}_j\ran \lan s^{2}_i|s^{2}_j\ran\ .
\end{equation}
Writing this out explicitly, we have
\bea
\lan s_1|s_4\ran =\lan\ua^1|\da^1\ran \lan\ua^2|\da^2\ran=1\ , \quad &&
\lan s_2|s_3\ran =\lan\ua^1|\da^1\ran \lan\da^2|\ua^2\ran=1\ , \nonumber \\
\lan s_3|s_2\ran =\lan\da^1|\ua^1\ran \lan\ua^2|\da^2\ran=1\ , \quad &&
\lan s_4|s_1\ran =\lan\da^1|\ua^1\ran \lan\da^2|\ua^2\ran=1\ .
\eea
The other inner products vanish.\
Based on these inner products for the basis states, we can write the
norm for the generic state as
\be\label{Eg.stateO(N)}
|\psi\ran = c_i|s_i\ran \quad\Rightarrow\quad \lan\psi|\psi\ran =
\big( c_j^* \lan s_j| \big)\cdot \big( c_i|s_i\ran \big) 
= c_1^*c_4 + c_4^*c_1 + c_2^*c_3 + c_3^*c_2\ .
\ee
Appropriate pairs of states can be used to define a new basis of
positive and negative norm states: $s_1\pm s_4$,\ $s_2\pm s_3$ and so
on have norm $\pm 2$ respectively.\
Likewise for $N=4$ flavours, the
$2^4=16$ configurations $|s_i\ran$ are \footnote{These states can
  be written in terms of a basis which is manifestly $O(N)$
  invariant.  Namely
  \begin{equation}
    \nonumber
    \begin{split}
      &|t_0\rangle = \prod_{i=1}^N |\uparrow^i\rangle\ ,\quad
      |t_{vec}\rangle = \prod_{j\not=i,i=1}^N
      |\uparrow^i\rangle|\downarrow^j\rangle\ , \quad \forall j\ ,\\
      & |t_{adj}\rangle = \prod_{j\not=k\not=i,i=1}^N
      |\uparrow^i\rangle|\downarrow^j\rangle\ , \quad \forall j, k\ ; \cdots
    \end{split}
  \end{equation}
  Where, $0$ is the trivial representation, $vec$ corresponds to the
  vector representation and $adj$ is the adjoint(antisymmetric)
  representation of $O(N)$.  To see the relation between the
  $|t\rangle$ and the $|s\rangle$ bases, consider for example, the
  states in \eqref{statesO(N)-siN=4}: these get organised into
  representations of $O(4)$ as {\bf 1}, {\bf 4}, {\bf 6}, {\bf 4},
  {\bf 1} with the states $|s_1\rangle$ and $|s_{16}\rangle$ forming
  two singlets, $|s_2\rangle$ to $|s_5\rangle$ ($|s_{12}\rangle$ to
  $|s_{15}\rangle$) belonging to the vector representation and
  $|s_6\rangle$ to $|s_{11}\rangle$ to the adjoint representations of
  $O(4)$.  We will, however, continue using the basis given in
  \eqref{basisStsiON} for convenience.}
\begin{equation}\label{statesO(N)-siN=4}
|s_1\ran=\left| {\tiny \begin{array}{c} \ua^1 \\ \ua^2 \\ \ua^3 \\ \ua^4
\end{array}} \right\ran , \quad
|s_2\ran=\left| {\tiny \begin{array}{c}	\ua^1 \\ \ua^2 \\ \ua^3 \\ \da^4
\end{array}} \right\ran , \quad
|s_3\ran=\left| {\tiny \begin{array}{c} \ua^1 \\ \ua^2 \\ \da^3 \\ \ua^4
\end{array}} \right\ran , \quad
\cdots\ , \quad
|s_{15}\ran=\left| {\tiny \begin{array}{c} \da^1 \\ \da^2 \\ \da^3 \\ \ua^4
\end{array}} \right\ran , \quad
|s_{16}\ran=\left| {\tiny \begin{array}{c} \da^1 \\ \da^2 \\ \da^3 \\ \da^4
\end{array}} \right\ran
\end{equation}
and the inner product \eqref{flavouredinnerproduct} is
\begin{equation}\label{flavouredinnerproductN=4}
\lan s_i|s_j\ran = \frac{1}{4!}\sum \epsilon_{A_1 A_2 A_3 A_4}\epsilon_{B_1 B_2 B_3 B_4}\lan s^{A_1}_i|s^{B_1}_j\ran \lan s^{A_2}_i|s^{B_2}_j\ran \lan s^{A_3}_i|s^{B_3}_j\ran \lan s^{A_4}_i|s^{B_4}_j\ran\ .
\end{equation}
Using \eqref{O(N)elementalproducts}, this simplifies to
\begin{eqnarray}
\lan s_i|s_j\ran &=& \frac{1}{4!}\sum_{A_1, A_2, A_3, A_4} \epsilon^2_{A_1 A_2 A_3 A_4}\lan s^{A_1}_i|s^{A_1}_j\ran \lan s^{A_2}_i|s^{A_2}_j\ran \lan s^{A_3}_i|s^{A_3}_j\ran \lan s^{A_4}_i|s^{A_4}_j\ran \nonumber \\
&=& \lan s^{1}_i|s^{1}_j\ran \lan s^{2}_i|s^{2}_j\ran \lan s^{3}_i|s^{3}_j\ran \lan s^{4}_i|s^{4}_j\ran\ ,
\end{eqnarray}
noting that there are $4!$ non-zero $\epsilon_{A_1 A_2 A_3 A_4}$
components, which can succinctly be written as
\begin{equation}
  \label{eq:1}
  \langle s_i|s_{17-i}\rangle=1\ ,
\end{equation}
with other inner products vanishing. Similarly, for general $N$, the inner product \eqref{flavouredinnerproduct}
is
\begin{equation}
\lan s_i|s_j\ran = \frac{1}{N!}\sum_{A_1, \dots, A_N} \epsilon^2_{A_1 \dots A_N}\lan s^{A_1}_i|s^{A_1}_j\ran \cdots \lan s^{A_N}_i|s^{A_N}_j\ran =\lan s^{1}_i|s^{1}_j\ran \cdots \lan s^{N}_i|s^{N}_j\ran =\prod_A \lan s^{A}_i|s^{A}_j\ran\ .
\end{equation}
Thus two generic states have inner product
\be
\lan\psi_1|\psi_2\ran = (\psi_1^{s_j})^* \psi_2^{s_i}\, \lan s_j| s_i\ran\, = (\psi_1^{s_j})^* \psi_2^{s_i}\, \prod_A \lan s_j^A| s_i^A\ran\,
\ee



\subsection{Ghost-spin chain for the $bc$-CFT with $O(N)$ symmetry}
\label{sec:gschainO(N)}

We want to now study infinite ghost-spin chains along the lines of
those in \cite{Jatkar:2017jwz}, but with additional flavour structure
respecting the $O(N)$ flavour symmetry we have been discussing so far.
In effect, this amounts to $N$ flavour copies of ghost-spins at each
lattice site. Thus we define two species of $N$-component commuting
spin variables, $\sigma^A_{bn}$ and $\sigma^A_{cn}$ at each lattice
site $n$ satisfying
\begin{equation}
\{\sigma^A_{bn},\sigma^B_{cn}\}=\delta^{AB}\ , \qquad [\sigma^A_{bn},\sigma^B_{bn'}]=[\sigma^A_{cn},\sigma^B_{cn'}]=[\sigma^A_{bn},\sigma^B_{cn'}]=0\ .
\end{equation}
These commuting spin-variables are Hermitian $\sigma_b^{\dagger A}=\sigma^A_b$, $\ \sigma_{c}^{\dagger A}=\sigma^A_c$ 
and their action on ghost-spin states is
\begin{equation}
\sigma^A_b|\downarrow^B\rangle=0, \quad \sigma^A_b|\uparrow^B\rangle=\delta^{AB}|\downarrow^B\rangle, \quad \sigma^A_c|\downarrow^B\rangle=\delta^{AB}|\uparrow^B\rangle, \quad \sigma^A_c|\uparrow^B\rangle=0,
\end{equation}
where there is no summation over the flavour index $B$. For multiple
ghost-spin states in the chain, the $\sigma_{bn}^A$ operator acts to
lower the $A^{th}$-flavour state within the ghost-spin configuration at
site $n$, and likewise $\sigma_{cn}^A$ is the corresponding raising
operator. To illustrate this explicitly, consider two ghost-spins with
$N=2$ flavours: then
\be
\sigma_{b1}^2 \big|^{\ua^1}_{\ua^2}{\big\ran}\big|^{\da^1}_{\da^2}{\big\ran}
= \big|^{\ua^1}_{\da^2}{\big\ran}\big|^{\da^1}_{\da^2}{\big\ran}\ ,\qquad
\sigma_{c2}^1 \big|^{\ua^1}_{\ua^2}{\big\ran}\big|^{\da^1}_{\da^2}{\big\ran}
= \big|^{\ua^1}_{\ua^2}{\big\ran}\big|^{\ua^1}_{\da^2}{\big\ran}\ ,\qquad
\sigma_{b1}^2 \sigma_{c2}^2
\big|^{\ua^1}_{\ua^2}{\big\ran}\big|^{\da^1}_{\da^2}{\big\ran}
= \big|^{\ua^1}_{\da^2}{\big\ran}\big|^{\da^1}_{\ua^2}{\big\ran}\ ,\quad \ldots
\ee


The ghost-spin chain is then defined by a Hamiltonian encoding
interactions between the ghost-spins at various lattice sites.
Since the flavours do not mix, the interaction Hamiltonian here is simply
a straightforward generalization involving a decoupled sum over various
flavours of the single flavour one in \cite{Jatkar:2017jwz}. So consider
a 1-dim ghost-spin chain with a ``hopping'' type interaction Hamiltonian
\begin{equation}\label{HforONgssigma}
H=J\sum_{n}\sum_{A=1}^{N}(\sigma^A_{bn}\sigma^A_{c(n+1)} +
\sigma^A_{bn}\sigma^A_{c(n-1)}) = J\sum_{n}\sum_{A=1}^{N}
(\sigma^A_{bn}\sigma^A_{c(n+1)}+\sigma^A_{b(n+1)}\sigma^A_{cn})\ ,
\end{equation}
where $n$, $n+1$, $n-1$ label nearest label lattice sites in the chain.
The action on a nearest neighbour pair of ghost-spins at lattice sites
$(n,n+1)$ is given as
\bea
\sigma^A_{bn}\sigma^A_{c(n+1)} : \left( \ldots\otimes \Big|\! {\tiny \bA{c}
  \vdots\\ \ua^A\\  \vdots \eA} \!\Big\ran_n\otimes
\Big|\! {\tiny \bA{c} \vdots\\ \da^A\\  \vdots \eA}
\!\Big\ran_{n+1}\otimes\ldots
\right)\!   &\longrightarrow& \!
J \left( \ldots\otimes \Big|\! {\tiny \bA{c}
  \vdots\\ \da^A\\  \vdots \eA} \!\Big\ran_n\otimes
\Big|\! {\tiny \bA{c} \vdots\\ \ua^A\\  \vdots \eA}
\!\Big\ran_{n+1}\otimes\ldots
\right) , \nonumber\\
\sigma^A_{b(n+1)}\sigma^A_{cn} : \left( \ldots\otimes \Big|\! {\tiny \bA{c}
  \vdots\\ \da^A\!\\  \vdots \eA} \!\Big\ran_n\otimes
\Big|\! {\tiny \bA{c} \vdots\\ \ua^A\\  \vdots \eA}
\!\Big\ran_{n+1}\otimes\ldots
\right)\!  &\longrightarrow& \!
J \left( \ldots\otimes \Big|\! {\tiny \bA{c}
  \vdots\\ \ua^A\\  \vdots \eA} \!\Big\ran_n\otimes
\Big|\! {\tiny \bA{c} \vdots\\ \da^A\\  \vdots \eA}
\!\Big\ran_{n+1}\otimes\ldots
\right) .\qquad
\eea


The $N$ flavours are decoupled and the interaction between ghost-spins
at two neighbouring lattice sites is through the same flavour at the
two sites. Thus we can follow the analysis in \cite{Jatkar:2017jwz}
flavour-by-flavour.

Towards constructing the continuum limit of the ghost-spin chain
(\ref{HforONgssigma}), we note that the $\sigma_{b,c}^A$ operators
commute at neighbouring lattice sites as in the single flavour case. 
Thus we define two species of $N$-component fermionic operators
satisfying the anti-commutation relations
\begin{equation}
\{a^A_{bi},a^B_{cj}\}=\delta_{ij}\delta^{AB}\ , \qquad \{a^A_{bi},a^B_{bj}\}=\{a^A_{ci},a^B_{cj}\}=0\ ,
\end{equation}
which anti-commute at different lattice sites $i,j$ also. The action of these fermionic operators on ghost-spin states is
\begin{eqnarray}
  && a^A_b|\downarrow^B\rangle=0\ , \quad a^A_b|\uparrow^B\rangle=\delta^{AB}|\downarrow^B\rangle\ , \quad a^A_c|\downarrow^B\rangle=\delta^{AB}|\uparrow^B\rangle\ , \quad a^A_c|\uparrow^B\rangle=0\ , \nonumber \\
  && \langle\downarrow^B|a^A_b=0\ , \quad \langle\uparrow^B|a^A_b=\delta^{AB}\langle\downarrow^B| \ , \quad \langle\downarrow^B|a^A_c=\delta^{AB}\langle\uparrow^B|\ , \quad \langle\uparrow^B|a^A_c=0\ .
\end{eqnarray}

This is obtained by constructing a flavoured generalization of the
Jordan-Wigner transformation in \cite{Jatkar:2017jwz} for the
commuting spin variables ($\sigma^A_b$, $\sigma^A_c$) as
\begin{eqnarray}\label{JWforsigma}
	&& \sigma^A_{b1}=a^A_{b1}\ , \quad \sigma^A_{c1}=a^A_{c1}\ , \qquad \sigma^A_{b2}=i(1-2a^A_{c1}a^A_{b1})a^A_{b2}\ , \quad \sigma^A_{c2}=-i(1-2a^A_{c1}a^A_{b1})a^A_{c2}\ , \quad \dots\ , \nonumber\\
	&& \sigma^A_{bn}=i(1-2a^A_{c1}a^A_{b1})i(1-2a^A_{c2}a^A_{b2})\dots i(1-2a^A_{c(n-1)}a^A_{b(n-1)})a^A_{bn}\ , \nonumber\\
	&& \sigma^A_{cn}=(-i)(1-2a^A_{c1}a^A_{b1})(-i)(1-2a^A_{c2}a^A_{b2})\dots (-i)(1-2a^A_{c(n-1)}a^A_{b(n-1)})a^A_{cn}\ , \quad \dots
\end{eqnarray}
for each flavour index $A$ independently (\ie\ the transformations are
decoupled for distinct flavours).
The inverse transformations for the fermionic ghost-spin variables
($a^A_b$, $a^A_c$) are
\begin{eqnarray}\label{iJWfora}
  && a^A_{b1}=\sigma^A_{b1}\ , \quad a^A_{c1}=\sigma^A_{c1}\ , \qquad a^A_{b2}=i(1-2\sigma^A_{c1}\sigma^A_{b1})\sigma^A_{b2}\ , \quad a^A_{c2}=-i(1-2\sigma^A_{c1}\sigma^A_{b1})\sigma^A_{c2}\ , \quad \dots\ , \nonumber\\
  && a^A_{bn}=i(1-2\sigma^A_{c1}\sigma^A_{b1})i(1-2\sigma^A_{c2}\sigma^A_{b2})\dots i(1-2\sigma^A_{c(n-1)}\sigma^A_{b(n-1)})\sigma^A_{bn}\ , \nonumber\\
  && a^A_{cn}=(-i)(1-2\sigma^A_{c1}\sigma^A_{b1})(-i)(1-2\sigma^A_{c2}\sigma^A_{b2})\dots (-i)(1-2\sigma^A_{c(n-1)}\sigma^A_{b(n-1)})\sigma^A_{cn}\ , \quad \dots
\end{eqnarray}
for each flavour index $A$ independently. The factor $(1-2\sigma^A_{ci}\sigma^A_{bi})$ is $-1$ or $+1$ depending on whether the $i$-th location is occupied by ($\uparrow^A$) or not ($\downarrow^A$), which means $(1-2\sigma^A_{ci}\sigma^A_{bi})^2=1$. Using
\begin{equation}
	[\pm i(1-2\sigma^A_{ci}\sigma^A_{bi})]^{\dagger}=\pm i(1-2\sigma^A_{ci}\sigma^A_{bi})\ ,
\end{equation}
we can check that the operators $a^A_{bn}$, $a^A_{cn}$ are hermitian. Now
substituting the Jordan-Wigner transformation \eqref{JWforsigma} in the
ghost-spin Hamiltonian in the commuting spin variables
\eqref{HforONgssigma} gives
\begin{eqnarray}
	H&=&J\sum_{n}\sum_{A=1}^{N}(\sigma^A_{bn}\sigma^A_{c(n+1)}+\sigma^A_{bn}\sigma^A_{c(n-1)})\ , \nonumber\\
	&=& J\sum_{n}\sum_{A=1}^{N}\left(i^{n-1}[1]^A[2]^A\dots [n-1]^A a^A_{bn}(-i)^n[1]^A[2]^A\dots [n]^A a^A_{c(n+1)}\right. \nonumber\\
	&&\qquad\qquad \left.+i^{n-1}[1]^A[2]^A\dots [n-1]^A a^A_{bn}(-i)^{n-2}[1]^A[2]^A\dots [n-2]^A a^A_{c(n-1)} \right)\ ,
\end{eqnarray}
where $[k]^A=(1-2a^A_{ck}a^A_{bk})$. Commuting the various $[k]^A$ factors gives
\begin{eqnarray}
  H&=&J\sum_{n}\sum_{A=1}^{N}\left((-i)a^A_{bn}(1-2a^A_{cn}a^A_{bn})a^A_{c(n+1)}+i(1-2a^A_{c(n-1)}a^A_{b(n-1)}) a^A_{bn}a^A_{c(n-1)} \right)\ , \nonumber\\
  &=& iJ\sum_{n}\sum_{A=1}^{N}a^A_{bn}(a^A_{c(n+1)}-a^A_{c(n-1)})\ .
\end{eqnarray}
We see that this Hamiltonian for the $1$-dimensional chain of
$N$-level ghost-spins with $O(N)$ symmetry breaks up as a decoupled
sum of $N$ copies of the Hamiltonian for 2-level ghost-spins 
in \cite{Jatkar:2017jwz}. Then following the analysis there 
for each flavour independently and taking the continuum limit, we can show
that we obtain $N$ copies of $bc$-ghost CFTs with $O(N)$ flavour
symmetry.\ 
This finally gives
\be
H = \sum_{A=1}^N \sum_k k b^A_{-k} c^A_k
= \sum_{A=1}^N \sum_{k>0} k \left( b^A_{-k} c^A_k + c_{-k}^A b_k^A \right)\ ,
\ee
which is essentially the operator $L_0$ for a $bc$-ghost CFT enjoying
$O(N)$-flavour symmetry, with action
\be
S = \int d^2z\, \sum_{A=1}^N\, b^A\del c^A\ ,
\ee
and a corresponding anti-holomorphic part. Further details are similar
to \cite{Jatkar:2017jwz}, except with multiple flavours.

\subsection{Correlated ghost-spin states and entanglement}\label{sec:CorrStO(N)}

We now return to ghost-spin ensembles and their entanglement
properties.  Along the lines in (\ref{basisStsiON}) for enumerating
states in the $\ua,\da$-basis, we can clearly use the $|\pm^i\ran$-basis
to define the basis states $|s_i\ran$ there: the advantage in the
$|\pm^i\ran$-basis is that positive/negative norm states are easier to
identify manifestly. For a single ghost-spin with $N=2$ flavours, we have
then
\be\label{statesO(N)-siN=2+-}
|s_1\ran = \big|^{+^1}_{+^2}{\big\ran}\ ,\qquad
|s_2\ran = \big|^{+^1}_{-^2}{\big\ran}\ ,\qquad
|s_3\ran = \big|^{-^1}_{+^2}{\big\ran}\ ,\qquad
|s_4\ran = \big|^{-^1}_{-^2}{\big\ran}\ .
\ee
We remind the reader that although we are using column vectors for
notational convenience, these are really direct product states:
the inner product \eqref{flavouredinnerproduct} here gives
\bea\label{N=2si.sj+-}
\lan s_1|s_1\ran = \lan +^1|+^1\ran \lan +^2|+^2\ran = 1 ,\quad &&
\lan s_4|s_4\ran = \lan -^1|-^1\ran \lan -^2|-^2\ran = 1 ,\ \nonumber\\
\lan s_2|s_2\ran = \lan +^1|+^1\ran \lan -^2|-^2\ran = -1 ,\quad &&
\lan s_3|s_3\ran = \lan -^1|-^1\ran \lan +^2|+^2\ran = -1 ,\ \
\eea
and this is an orthonormal basis of positive and negative norm states.
A generic state then has norm
\be
|\psi\ran = c_i|s_i\ran\quad \Rightarrow\quad
\lan\psi|\psi\ran = |c_1|^2 + |c_4|^2 - |c_2|^2 - |c_3|^2\ .
\ee
Thus states made from $|s_2\ran,\, |s_3\ran$ alone have negative norm.
It is straightforward to write down similar basis states for arbitrary
$N$ flavours. For $N$ flavours, there are $2^N$ basis states $|s_i\ran$.
The inner products are
\be
\lan s_i|s_i\ran = \prod_A \lan s_i^A|s_i^A\ran\ .
\ee
Since this is a diagonal basis now, we have\ $\lan s_i|s_i\ran = \pm 1$
respectively when there is an even or odd number of $|-^A\ran$ elemental
ghost-spins in $|s_i\ran$. This is exemplified in the $N=2$ case
(\ref{N=2si.sj+-}) above.

Now let us consider two ghost-spins. The states can be made from the $2^{2N}$
basis states\ $|s_i\ran|s_j\ran$\ obtained by tensor products of the
single ghost-spin states. The inner products between them are
\be\label{2gs:sidotsj+-}
\lan (s_k|\lan s_l|\,)\cdot (\,|s_i\ran|s_j\ran) = \lan s_k|s_i\ran\,
\lan s_l|s_j\ran\,
\ee
Then the general state and its norm are
\be
|\psi\ran=\psi^{s_is_j}|s_i\ran |s_j\ran\ ,\qquad
\lan\psi|\psi\ran = (\psi^{s_ks_l})^* \psi^{s_is_j}\,
\prod_A \lan s_k^A|s_i^A\ran\, \prod_B \lan s_l^B|s_j^B\ran\,\ ,
\ee
with two products over the flavour components of the two basis states.
Tracing over say the second ghost-spin in this state leads to a subsystem
comprising the single remaining ghost-spin, with reduced density matrix
defined as
\be\label{RDM2gsN}
(\rho_A)^{s_k,s_i} = (\psi^{s_ks_l})^* \psi^{s_is_j}\,
\lan s_l|s_j\ran = (\psi^{s_ks_l})^* \psi^{s_is_j}\,
\prod_B \lan s_l^B|s_j^B\ran\ .
\ee
The entanglement entropy of this reduced density matrix can then be
calculated using the formulation in
\cite{Narayan:2016xwq,Jatkar:2016lzq,Jatkar:2017jwz}: we will see this
below.
Restricting attention for simplicity to $N=2$ flavours, we can use
the four basis states (\ref{statesO(N)-siN=2+-}). Then the 2 ghost-spin
states can be described using the 16 basis states
$|s_{i,j}\ran\equiv |s_i\ran|s_j\ran$, or more explicitly,
\bea
\big|^{+^1}_{+^2}{\big\ran}\big|^{+^1}_{+^2}{\big\ran} ,\ \
\big|^{+^1}_{+^2}{\big\ran}\big|^{+^1}_{-^2}{\big\ran} ,\ \
\big|^{+^1}_{+^2}{\big\ran}\big|^{-^1}_{+^2}{\big\ran} ,\ \
\big|^{+^1}_{+^2}{\big\ran}\big|^{-^1}_{-^2}{\big\ran} ,\ \
\big|^{+^1}_{-^2}{\big\ran}\big|^{+^1}_{+^2}{\big\ran} ,\ \
\big|^{+^1}_{-^2}{\big\ran}\big|^{+^1}_{-^2}{\big\ran} ,\ \
\big|^{+^1}_{-^2}{\big\ran}\big|^{-^1}_{+^2}{\big\ran} ,\ \
\big|^{+^1}_{-^2}{\big\ran}\big|^{-^1}_{-^2}{\big\ran} ,&& \nonumber\\
\big|^{-^1}_{+^2}{\big\ran}\big|^{+^1}_{+^2}{\big\ran} ,\ \
\big|^{-^1}_{+^2}{\big\ran}\big|^{+^1}_{-^2}{\big\ran} ,\ \
\big|^{-^1}_{+^2}{\big\ran}\big|^{-^1}_{+^2}{\big\ran} ,\ \
\big|^{-^1}_{+^2}{\big\ran}\big|^{-^1}_{-^2}{\big\ran} ,\ \
\big|^{-^1}_{-^2}{\big\ran}\big|^{+^1}_{+^2}{\big\ran} ,\ \
\big|^{-^1}_{-^2}{\big\ran}\big|^{+^1}_{-^2}{\big\ran} ,\ \
\big|^{-^1}_{-^2}{\big\ran}\big|^{-^1}_{+^2}{\big\ran} ,\ \
\big|^{-^1}_{-^2}{\big\ran}\big|^{-^1}_{-^2}{\big\ran} ,&&\ \ \
\eea
The inner products (\ref{2gs:sidotsj+-}) can then be seen to give the
norms \eg\
\be
\lan s_{i,i}|s_{i,i}\ran = (\lan s_i|s_i\ran)^2 = 1 ,\ \ 
\lan s_{1,2}|s_{1,2}\ran = \lan s_1|s_1\ran \lan s_2|s_2\ran = -1 ,\ \ 
\lan s_{2,3}|s_{2,3}\ran = \lan s_2|s_2\ran \lan s_3|s_3\ran = 1 ,\
\ldots
\ee
and so on, using (\ref{N=2si.sj+-}). It is clear again that the norms
are again $\pm 1$ depending on whether the state $|s_{i,j}\ran$ contains
an even or odd number of $|-^A\ran$ elemental states.
The general state has norm
\be\label{2gsGen}
|\psi\ran = \sum \psi^{s_i,s_j} |s_i\ran |s_j\ran \qquad
\lan\psi|\psi\ran = (\psi^{s_k,s_l})^* \psi^{s_i,s_j} \lan s_k|s_i\ran
\lan s_l|s_j\ran = |\psi^{s_i,s_j}|^2 \lan s_{i,j}|s_{i,j}\ran\ .
\ee
This is a sum over $|\psi^{s_i,s_j}|^2$ weighted by $\pm 1$ depending
on the sign of the norm of $|s_{i,j}\ran$.
A particularly interesting subset of states are what we call
``correlated states'', generalizing the discussion in
\cite{Jatkar:2016lzq}.\ These are of the form
\be\label{2gsCorrStgen}
|\psi^{corr}\ran = \sum \psi^{s_i,s_i} |s_i\ran |s_i\ran \qquad
\lan\psi^{corr}|\psi^{corr}\ran = \sum_{i=1}^4 |\psi^{s_j,s_j}|^2\ > 0\ ,
\ee
and are necessarily positive norm, even though some of the individual
basis states are negative norm. The basis states $|s_i\ran|s_i\ran$
here are of the form
\be
\big|^{+^1}_{+^2}{\big\ran}\big|^{+^1}_{+^2}{\big\ran} ,\quad
\big|^{+^1}_{-^2}{\big\ran}\big|^{+^1}_{-^2}{\big\ran} ,\quad
\big|^{-^1}_{+^2}{\big\ran}\big|^{-^1}_{+^2}{\big\ran} ,\quad
\big|^{-^1}_{-^2}{\big\ran}\big|^{-^1}_{-^2}{\big\ran} ,
\ee
and we see explicitly that this subspace of correlated states is obtained
by entangling some configuration for the first ghost-spin with an identical
configuration for the second ghost-spin. Thus we have only 4 states which
span the correlated ghost-spin subspace. It is clear that these are
necessarily positive norm since there is an even number of minus
ghost-spins (any odd number in each column is doubled). Note that
this is a smaller subspace than that comprising all positive norm states
which simply need to have an even number of minus signs: \eg\ the
basis state $\big|^{+^1}_{+^2}{\big\ran}\big|^{-^1}_{-^2}{\big\ran}$
is positive norm but the two ghost-spins have different configurations.
This can be generalized to two ghost-spins with $N$ flavours in a
straightforward manner: the general state is again of the form
(\ref{2gsCorrStgen}) but with the $|s_i\ran$ encoding $N$ flavour
ghost-spin configurations.
There are $2^N$ basis states $|s_i\ran$ so this subspace of correlated
states is $2^N$-dimensional, somewhat smaller than the
$2^{2N}$-dimensional space of all states.

These correlated ghost-spin states entangle identical ghost-spins
between the two sets of ghost-spins.  These states necessarily encode
positive entanglement since any sublinear combination of the norm is
still positive definite (one way to see this is to note that this can
be mapped to an auxiliary system of ordinary spins, which has no minus
signs and is entirely positive norm). More explicitly, for a state of
the form
$|\psi\ran = \psi^{s_I,s_I}|s_I\ran|s_I\ran + \psi^{s_J,s_J}|s_J\ran|s_J\ran$
made of the states $|s_{I,I}\ran, |s_{J,J}\ran$,\ 
the reduced density matrix (\ref{RDM2gsN}) can be taken to construct
a mixed-index reduced density matrix as in
\cite{Narayan:2016xwq,Jatkar:2016lzq,Jatkar:2017jwz}, which then makes
explicit the contraction structure with respect to the ghost-spin
inner product metric (incorporating the signs for negative norm
states). To be explicit, consider\
$|\psi\ran = \psi^{s_1,s_1}|s_1\ran|s_1\ran + \psi^{s_2,s_2}|s_2\ran|s_2\ran$
noting that $|s_1\ran$ and $|s_2\ran$ are positive and negative norm
respectively. Then 
\be\label{RDM2gscorrO(N)}
(\rho_A)^{s_1}_{s_1} = (\rho_A)^{s_1,s_1} = |\psi^{s_1,s_1}|^2 ,\quad
(\rho_A)^{s_2}_{s_2} = -(\rho_A)^{s_2,s_2} = |\psi^{s_2,s_2}|^2\ ;\qquad
{\rm tr}\rho_A = \lan\psi|\psi\ran = 1\ .
\ee
The mixed-index reduced density matrix here is\
$(\rho_A)^{s_i}_{s_j} = \gamma_{s_j,s_k} (\rho_A)^{s_i,s_k} =
\lan s_j|s_k\ran (\rho_A)^{s_i,s_k}$\,,\
where the metric $\gamma_{s_i,s_j}=\lan s_i|s_j\ran$ is defined by the
inner products (\ref{N=2si.sj+-}).\
This description can be generalized to all such states, and indeed to
all 2-ghost-spin states (\ref{2gsGen}): in this case, it can be shown
along the lines of the single flavour case that more general positive
norm subsectors exist. In general however, the state space has many
branches of negative norm states, and the reduced density matrix in
general has negative eigenvalues with a complex entanglement entropy
correspondingly (as was already the case in the single flavour case).

These states can be generalized to any even number of ghost-spins.
For odd numbers of ghost-spins however, this structure does not
prevail: there are states that are positive norm but the reduced
density matrix continues to have negative eigenvalues so that the
entanglement entropy is not positive.

It is now interesting to consider two copies of ghost-spin chains and
consider correlated ghost-spin states representing entanglement
between the two chains. This is motivated by the discussion and
picture in \cite{Narayan:2017xca} of $dS_4$ as dual to a
thermofield-double type entangled state in two copies $CFT_F\times
CFT_P$ of the ghost-CFT at $I^+$ and $I^-$\ (reviewed in the Discussion
in sec.~\ref{sec:discussion}).  So let us consider
${\cal GC}_1\times {\cal GC}_2$ where each ${\cal GC}$ represents
a ghost-spin chain whose continuum limit gives a $bc$-ghost CFT with
flavour symmetry as in sec.~\ref{sec:gschainO(N)}.
Configurations of each ${\cal GC}$ can be represented schematically by
\be
|\sigma\ran\ \equiv\ (\ldots |s_n\ran |s_{n+1}\ran\ldots )
\ee
Then correlated entangled states in ${\cal GC}_1\times {\cal GC}_2$
can be represented as
\be\label{corrStgsc}
|\psi\ran = \psi^{\sigma,\sigma} |\sigma\ran|\sigma\ran\ ,\qquad
\lan\psi|\psi\ran = \sum_{|\sigma\ran} |\psi^{\sigma,\sigma}|^2 > 0\ .
\ee
Now the states $|\sigma\ran$ include the ground state as well as
excited states. If we restrict to the ground states alone, then 
since the flavours are all decoupled from each other, the ground
states $|\sigma\ran$ comprise a $2^N$-dimensional subspace noting 
that each $|s_n\ran$ at lattice site $n$ in $|\sigma\ran$ has
$2^N$ possibilities. Thus tracing over the second ghost-spin chain
copy ${\cal GC}_2$, we obtain the entanglement entropy of
$|\psi\ran$ restricting to ground states $|\sigma_g\ran$ as
\be\label{corrStEE}
S_A = -\sum_{i=1}^{2^N} |\psi^{\sigma_g,\sigma_g}|^2\,\log |\psi^{\sigma_g,\sigma_g}|^2
\quad\ra\quad -2^N\, {1\over 2^N}\,\log {1\over 2^N} = N\log 2\ .
\ee
We have used $\sum_{g} |\psi^{\sigma_g,\sigma_g}|^2 = 1$ from normalization
and imposed maximal entanglement, which equates all the coefficients
giving $|\psi^{\sigma_g,\sigma_g}|^2={1\over 2^N}$\,. Thus the entanglement
entropy scale as the number of flavours $N$.

\subsection{Symmetric, spin-glass type, inner products}\label{sec:symmetric-spin-glass}

Here we briefly mention a generalization of the $O(N)$ flavoured case
but with the various flavours talking to each other, with a spin glass
type coupling. We define the elemental inner products \eqref{NJAB}
using a symmetric form $J^{AB}$ and take the inner products between
the configurations $|s_i\ran$ to be \eqref{flavouredinnerproduct}.

Let us consider first $N=2$ flavours: then the states are as in
(\ref{statesO(N)-siN=2}) and the inner products are
\begin{equation}
  \lan s_i|s_j\ran = \lan s^{1}_i|s^{1}_j\ran \lan s^{2}_i|s^{2}_j\ran
  + \lan s^{1}_i|s^{2}_j\ran \lan s^{2}_i|s^{1}_j\ran\ .
\end{equation}
Using the elemental inner products \eqref{NJAB}, the non-zero inner
products are
\bea
&& \lan s_1|s_4\ran =\lan\ua^1|\da^1\ran \lan\ua^2|\da^2\ran + \lan\ua^1|\da^2\ran \lan\ua^2|\da^1\ran=J^{11}J^{22}+J^{12}J^{21}\ , \nonumber \\
&& \lan s_4|s_1\ran =\lan\da^1|\ua^1\ran \lan\da^2|\ua^2\ran + \lan\da^1|\ua^2\ran \lan\da^2|\ua^1\ran = J^{11}J^{22}+J^{12}J^{21}\ , \\
&& \lan s_2|s_2\ran =\lan\ua^1|\da^2\ran \lan\da^2|\ua^1\ran = J^{12}J^{21}\ , \qquad \lan s_2|s_3\ran =\lan\ua^1|\da^1\ran \lan\da^2|\ua^2\ran = J^{11}J^{22}\ , \nonumber \\
&& \lan s_3|s_2\ran =\lan\da^1|\ua^1\ran \lan\ua^2|\da^2\ran = J^{11}J^{22}\ , \qquad \lan s_3|s_3\ran =\lan\da^1|\ua^2\ran \lan\ua^2|\da^1\ran = J^{12}J^{21}\ . \nonumber
\eea
The metric in the space of $|s_i\ran$'s is real, symmetric and its
determinant is $(\det J) (J^{11}J^{22}+J^{12}J^{21})^3$, where
$\det\,J = J^{11}J^{22}-J^{12}J^{21}$. For an orthogonal matrix
$J^{AB}$, $\det J \neq 0$ and the metric is non-singular only if
$J^{11}J^{22}+J^{12}J^{21} \neq 0$.

We will not dwell more on this, although this may be worth investigating
further.

\section{Symplectic inner products}\label{sec:sympl}

We want to study ``symplectically flavoured'' ghost-spins. We introduce the symplectic structure by defining the elemental inner products with an anitsymmetric matrix\,:
\be\label{2NOmegaABwithi}
\langle\uparrow^A|\downarrow^B\rangle=i\,\Omega^{AB}\ , \quad \langle\downarrow^A|\uparrow^B\rangle=i\,\Omega^{AB}\ , \quad \langle\uparrow^A|\uparrow^B\rangle=0=\langle\downarrow^A|\downarrow^B\rangle\ ; \quad\ A,B=1,\dots,2N\ ,
\ee
where
$\Omega^{AB}$ is a symplectic form, which is antisymmetric, \ie\ \ $\Omega^{AB}=-\Omega^{BA}$. We will take the only nonzero elements as
\be\label{OmegaAB}
\Omega^{12}=1=-\Omega^{21} ,\qquad \Omega^{34}=1=-\Omega^{43} ,\qquad\dots ,
\qquad \Omega^{2N-1\, 2N}=1=-\Omega^{2N\, 2N-1}\ .
\ee
For a single symplectically flavoured ghost-spin there are $2^{2N}$
distinct configurations comprising the basis states $|s_1\ran$,
$\dots$, $|s_{2^{2N}}\ran$ and a generic state is
\be\label{symplNorm}
|\psi\ran = \psi^{s_i}|s_i\ran\quad\Rightarrow\quad
\lan\psi|\psi\ran = (\psi^{s_j})^*\, \psi^{s_i}\,
\lan s_j\,|\,s_i\ran\ .
\ee
We define inner products $\lan s_j\,|\,s_i\ran$ between the basis states as
\begin{equation}\label{2NOmegaAB-innerproduct}
\lan s_i|s_j\ran = \frac{1}{(2N)!}\sum \epsilon_{A_1 A_2 \dots A_{2N}}\epsilon_{B_1 B_2 \dots B_{2N}}\lan s^{A_1}_i|s^{B_1}_j\ran \lan s^{A_2}_i|s^{B_2}_j\ran \cdots \lan s^{A_{2N}}_i|s^{B_{2N}}_j\ran\ ,
\end{equation}
where $i,j=1,2,\dots,2^{2N}$ label the configurations, $A_1, B_1, \dots
=1,2,\dots,2N$ label the flavours and $\epsilon_{A_1 A_2 \dots A_{2N}}$ is
the totally symmetric tensor with $\epsilon_{123\,\dots\, 2N}=1$ and
$\epsilon_{A_1 A_2 \dots A_{2N}}$ vanishes if any two labels are the same.
Thus as in (\ref{flavouredinnerproduct}), $\epsilon_{A_1 A_2 \dots A_{2N}}$
ensures that each elemental state $|s_j^A\ran$ in $|s_j\ran$ is paired
with another corresponding elemental state in $\lan s_i|$.

Let us consider first a single ghost-spin with $2$ flavours ($N=1$):
then the distinct configurations comprise the four basis states
\be\label{statesSympl-siN=2}
|s_1\ran = \big|^{\ua^1}_{\ua^2}{\big\ran}\ ,\qquad
|s_2\ran = \big|^{\ua^1}_{\da^2}{\big\ran}\ ,\qquad
|s_3\ran = \big|^{\da^1}_{\ua^2}{\big\ran}\ ,\qquad
|s_4\ran = \big|^{\da^1}_{\da^2}{\big\ran}\ ,
\ee
\ie\ in $|s_1\ran$, the first flavour is $\ua^1$ and the second
flavour is $\ua^2$, and likewise for $|s_2\ran, |s_3\ran, |s_4\ran$. The
non-zero elemental inner products in \eqref{2NOmegaABwithi} are
\be\label{ua1da2etc}
\lan\ua^1|\da^2\ran = i = \lan\da^1|\ua^2\ran\ ,
\qquad \lan\ua^2|\da^1\ran = -i = \lan\da^2|\ua^1\ran
\ee
and the inners products \eqref{2NOmegaAB-innerproduct} between the
configurations simplify to
\begin{equation}
	\lan s_i|s_j\ran = \lan s^{1}_i|s^{2}_j\ran \lan s^{2}_i|s^{1}_j\ran\ .
\end{equation}
Then non-zero inner products for the configurations \eqref{statesO(N)-siN=2} are
\bea
&& \lan s_1|s_4\ran =\lan\ua^1|\da^2\ran \lan\ua^2|\da^1\ran=1\ , \quad
\lan s_4|s_1\ran =\lan\da^1|\ua^2\ran \lan\da^2|\ua^1\ran=1\ , \nonumber \\
&& \lan s_2|s_2\ran =\lan\ua^1|\da^2\ran \lan\da^2|\ua^1\ran=1\ , \quad
\lan s_3|s_3\ran =\lan\da^1|\ua^2\ran \lan\ua^2|\da^1\ran=1\ .
\eea
These give a real, symmetric and non-singular metric in the space of
configurations $|s_i\ran$.\
Based on these inner products, we can write the norm for the generic state as
\be\label{sympNormN=1}
|\psi\ran = c_i|s_i\ran :\qquad \lan\psi|\psi\ran
= c_1^*c_4+c_4^*c_1 + |c_2|^2 + |c_3|^2\ .
\ee

Likewise for a single ghost-spin with $4$ flavours ($N=2$), there are $16$ distinct configurations comprising of basis states \eqref{statesO(N)-siN=4}. The non-zero elemental inner products with $\Omega^{12}=1$ and $\Omega^{34}=1$ are
\begin{eqnarray}
	&& \lan\ua^1|\da^2\ran = i = \lan\da^1|\ua^2\ran\ ,
	\qquad \lan\ua^2|\da^1\ran = -i = \lan\da^2|\ua^1\ran\ , \nonumber \\
	&& \lan\ua^3|\da^4\ran = i = \lan\da^3|\ua^4\ran\ ,
	\qquad \lan\ua^4|\da^3\ran = -i = \lan\da^4|\ua^3\ran\ .
\end{eqnarray}
Since these are the only non-zero elemental inner products, the inner products \eqref{2NOmegaAB-innerproduct} for the basis states \eqref{statesO(N)-siN=4} reduce to
\begin{equation}
	\lan s_i|s_j\ran = \frac{1}{4!}\sum \epsilon_{A_1 A_2 A_3 A_4}\epsilon_{\tilde{A}_1 \tilde{A}_2 \tilde{A}_3 \tilde{A}_4}\lan s^{A_1}_i|s^{\tilde{A}_1}_j\ran \lan s^{A_2}_i|s^{\tilde{A}_2}_j\ran \lan s^{A_3}_i|s^{\tilde{A}_3}_j\ran \lan s^{A_4}_i|s^{\tilde{A}_4}_j\ran\ ,
\end{equation}
where $\tilde{A_k}=|\Omega^{\tilde{A}_k A_l}|A_l$, \eg\ for $A_1=1$, $\tilde{A}_1=2$, for $A_2=4$, $\tilde{A}_2=3$, etc. As there are only $4!$ such non-zero terms in the above inner product, it becomes
\be
\lan s_i|s_j\ran = \epsilon_{1234}\epsilon_{2143}\lan s^{1}_i|s^{2}_j\ran \lan s^{2}_i|s^{1}_j\ran \lan s^{3}_i|s^{4}_j\ran \lan s^{4}_i|s^{3}_j\ran = \lan s^{1}_i|s^{2}_j\ran \lan s^{2}_i|s^{1}_j\ran \lan s^{3}_i|s^{4}_j\ran \lan s^{4}_i|s^{3}_j\ran\ .
\ee
The non-zero inner products between $|s_i\ran$'s computed using this formula can be written compactly as
\begin{equation}\label{sympNormN=2}
	\lan s_i|\tilde{s}_j\ran = 1\ ,
\end{equation}
where $|\tilde{s}_j\ran$ is defined such that if the $A$-th flavour entry in $|s_i\ran$ is $\ua^A$ (or $\da^A\ran$) then the $\tilde{A}$-th flavour entry in $|\tilde{s}_j\ran$ is $\da^{\tilde{A}}$ (or $\ua^{\tilde{A}}\ran$) for $\tilde{A}=|\Omega^{\tilde{A}A}|A$. We see that the metric $\lan s_i|s_j\ran$ is real, symmetric and non-singular.

We can generalize this to $2N$ flavours, where the non-zero elemental inner products are
\begin{eqnarray}
	&& \lan\ua^1|\da^2\ran = i\ , \qquad \lan\ua^3|\da^4\ran = i\ ,\quad\qquad \dots\dots\ , \qquad \lan\ua^{2N-1}|\da^{2N}\ran = i\ , \nonumber \\
	&& \lan\da^1|\ua^2\ran =i\ , \qquad \lan\da^3|\ua^4\ran = i\ , \quad\qquad \dots\dots\ , \qquad \lan\da^{2N-1}|\ua^{2N}\ran = i\ , \nonumber \\
	&& \lan\ua^2|\da^1\ran = -i\ , \quad \lan\ua^4|\da^3\ran = -i\ , \ \qquad \dots\dots\ , \qquad \lan\ua^{2N}|\da^{2N-1}\ran = -i\ , \nonumber \\
	&& \lan\da^2|\ua^1\ran = -i\ , \quad \lan\da^4|\ua^3\ran = -i\ , \ \qquad \dots\dots\ , \qquad \lan\da^{2N}|\ua^{2N-1}\ran = -i\ .
\end{eqnarray}
Then the inner products \eqref{2NOmegaAB-innerproduct} become
\begin{eqnarray}
	\lan s_i|s_j\ran &=& \frac{1}{(2N)!}\sum \epsilon_{A_1 A_2 \dots A_{2N}}\epsilon_{\tilde{A}_1 \tilde{A}_2 \dots \tilde{A}_{2N}}\lan s^{A_1}_i|s^{\tilde{A}_1}_j\ran \lan s^{A_2}_i|s^{\tilde{A}_2}_j\ran \cdots \lan s^{A_{2N}}_i|s^{\tilde{A}_{2N}}_j\ran\ , \nonumber \\
	&=& \epsilon_{1234\dots 2N-1\, 2N}\epsilon_{2143\dots 2N\, 2N-1} \lan s^{1}_i|s^{2}_j\ran \lan s^{2}_i|s^{1}_j\ran \cdots \lan s^{2N-1}_i|s^{2N}_j\ran \lan s^{2N}_i|s^{2N-1}_j\ran\ , \nonumber \\
	&=& \lan s^{1}_i|s^{2}_j\ran \lan s^{2}_i|s^{1}_j\ran \cdots \lan s^{2N-1}_i|s^{2N}_j\ran \lan s^{2N}_i|s^{2N-1}_j\ran\ ,
\end{eqnarray}
where $\tilde{A_k}=|\Omega^{\tilde{A}_k A_l}|A_l$. Using the elemental inner products, we see that the non-zero inner products are $\lan s_i|\tilde{s}_{j}\ran=1$, with $|\tilde{s_j}\ran$ as defined earlier.

Thus the norm of a generic state $|\psi\ran = \psi^{s_i}|s_i\ran$ is
\be\label{symplNorm2N}
\lan\psi|\psi\ran = (\psi^{s_j})^*\, \psi^{s_i}\,
\lan s_j\,|\,s_i\ran = (\psi^{s_j})^*\, \psi^{s_i} \lan s^{1}_j|s^{2}_i\ran \lan s^{2}_j|s^{1}_i\ran \cdots \lan s^{2N-1}_j|s^{2N}_i\ran \lan s^{2N}_j|s^{2N-1}_i\ran\ .
\ee
Along the same lines, the general inner product between any two states is
\be
\lan\psi_1|\psi_2\ran = (\psi_1^{s_j})^*\, \psi_2^{s_i}\,
\lan s_j\,|\,s_i\ran = (\psi_1^{s_j})^*\, \psi_2^{s_i} \lan s^{1}_j|s^{2}_i\ran \lan s^{2}_j|s^{1}_i\ran \cdots \lan s^{2N-1}_j|s^{2N}_i\ran \lan s^{2N}_j|s^{2N-1}_i\ran\ .
\ee

\medskip

\noindent {\bf Correlated ghost-spin states:}\ \
We want to now construct correlated ghost-spin states that are positive
norm and positive entanglement, along the lines of the discussion for
$O(N)$ flavoured cases in sec.~\ref{sec:CorrStO(N)}. This is most
transparent in a diagonal basis where positive and negative norm states
are manifest. For concreteness, let us consider the basis states
(\ref{statesSympl-siN=2}) for a single ghost-spin with 2 flavours, \ie\
$N=1$. The norm (\ref{sympNormN=1}) for a generic state can be recast
using a diagonal basis $|s_{\pm}\ran, |s_2\ran, |s_3\ran$ as
\be
|s_{\pm}\ran={1\over\sqrt{2}}(|s_1\ran\pm |s_4\ran):\qquad
\lan\psi|\psi\ran = |c_2|^2 + |c_3|^2 + |c_+|^2 - |c_-|^2\ .
\ee
Thus there are 3 basis states with positive norm and one with negative
norm. This can be carried out for more flavours as well. For $N=2$ for
instance, we have 16 basis states which can be recast as 10 positive
norm and 6 negative norm states, using (\ref{sympNormN=2}): besides
the states with $\lan s_i|s_i\ran\neq 0$, there are states with
off-diagonal inner products like $|s_{1,4}\ran$ above whose linear
combinations then add to the set of diagonal positive
norm
states. Note that the numbers of positive and negative norm basis
states are not equal.  With general $N$, \ie\ $A,B=1,\ldots,2N$, it
can be seen that there are $2^{2N}$ basis states in all: of these
there are\ ${2^{2N}+2^N\over 2}$ positive norm states
and ${2^{2N}-2^N\over 2}$ negative norm states
(this is easily verified for
$N=1,2$ above). For large $N$, we see that the number of positive
and negative norm states become asymptotically equal.

In terms of such a diagonal basis, we can consider 2 ghost-spins and
explicitly construct correlated ghost-spin states similar structurally
to (\ref{2gsCorrStgen}) in the $O(N)$ flavoured case. Let us label
these diagonal basis states for a single ghost-spin as $|s_i\ran$
(which should not be confused with the earlier nondiagonal $|s_i\ran$
basis).
Then the 2-ghost-spin states can be made from the $2^{4N}$ basis
states\ $|s_i\ran|s_j\ran$\ obtained by tensor products of the
single ghost-spin states. The general state and its norm are then
similar in structure to (\ref{2gsGen}) for the $O(N)$ case
sec.~\ref{sec:CorrStO(N)}. Correlated ghost-spin states can then be
constructed as in (\ref{2gsCorrStgen}) giving\ 
$|\psi^{corr}\ran=\sum\psi^{s_i,s_i}|s_i\ran|s_i\ran$\,: it can be
seen that these are positive norm and positive entanglement as in
(\ref{2gsCorrStgen}), (\ref{RDM2gscorrO(N)}).
This subspace has dimension $2^{2N}$, the number of basis states.
Since the details here are very similar to that in
sec.~\ref{sec:CorrStO(N)}, we will not describe them further here.

It is worth noting that the symplectic invariance is at the level of
the elemental ghost-spin basis states $\{|\ua^A\ran,\ |\da^A\ran\}$\,:
generic basis states
$|v_i\ran = v_i^{\ua^A}|\ua^A\ran + v_i^{\da^A}|\da^A\ran$,
have inner product
\begin{equation}
\lan v_1|v_2\ran = (v_1^{\ua^A})^*v_2^{\da^B}\lan \ua^A|\da^B\ran + (v_1^{\da^A})^*v_2^{\ua^B}\lan \da^A|\ua^B\ran = i[ (v_1^{\ua^A})^*v_2^{\da^B}\Omega^{AB} + (v_1^{\da^A})^*v_2^{\ua^B}\Omega^{AB} ]\ ,
\end{equation}
which is invariant under symplectic transformations. To see this
explicitly, consider two flavours ($N=1$) for simplicity.
Then a symplectic transformation by a real pseudo-orthogonal matrix
$R\in Sp(2)$ is\ $R^T \mathbf{\Omega} R = \mathbf{\Omega} , \
R\mathbf{\Omega} R^T = \mathbf{\Omega} , \
R^{-1}=-\mathbf{\Omega} R^T \mathbf{\Omega}$\,,
where $\mathbf{\Omega}$ is the symplectic form with
$\Omega^{12}=1=-\Omega^{21}$, and $\Omega^{-1}=-\Omega$. Thus
$(v_1^{\ua^A})^*v_2^{\da^B}\Omega^{AB}$ and $(v_1^{\da^A})^*v_2^{\ua^B}\Omega^{AB}$
are invariant under
$|\ua^A\ran\ra R^{AB}|\ua^B\ran,\ |\da^A\ran\ra R^{AB}|\da^B\ran$, \ie\
\begin{eqnarray}
&& (\tilde{v}_1^{\ua^A})^*\Omega^{AB}\tilde{v}_2^{\da^B} = (v_1^{\ua^C})^*R^{CA}\Omega^{AB} R^{BD}v_2^{\da^D} = (v_1^{\ua^C})^* \Omega^{CD} v_2^{\da^D}\ , \nonumber \\
&& (\tilde{v}_1^{\da^A})^*\Omega^{AB}\tilde{v}_2^{\ua^B} = (v_1^{\da^C})^*R^{CA}\Omega^{AB} R^{BD}v_2^{\ua^D} = (v_1^{\da^C})^* \Omega^{CD} v_2^{\ua^D}\ ,
\end{eqnarray}
where we have used
$R^{CA}\Omega^{AB} R^{BD} = R^T\mathbf{\Omega} R = \mathbf{\Omega}$.

The elemental inner products (\ref{ua1da2etc}) are consistent with (and
motivated by) an operator algebra alongwith states, defined as\ (these
arise in theories of symplectic fermions \cite{Kausch:1995py})
\be
\{ \sigma_b^A, \sigma_c^B \} = i\,\Omega^{AB}\hat{K}\ ;\qquad
|\ua^A\ran = \sigma_c^A|\da\ran\ ,\quad
\lan\da^A| = \lan\ua|\sigma_b^A\ ,\quad
|\da^A\ran = \sigma_b^A|\ua\ran\ ,\quad
\lan\ua^A| = \lan\da|\sigma_c^A\ ,
\ee
where $|\ua\ran$ and $|\da\ran$ are ghost-spin states with\
$\lan\ua|\da\ran=1=\lan\da|\ua\ran$. The hermiticity of $\{ \sigma_b^A, \sigma_c^B \}$ for hermitian $\sigma_b^A$, $\sigma_c^B$ and real $\Omega^{AB}$ gives $\hat{K}^{\dagger}=-\hat{K}$ \ie\ $\hat{K}$ is anti-hermitian. This anti-Hermitian operator leads
\begin{equation}
	(\lan \ua|\hat{K}|\da\ran)^{\dagger} = \lan \da|\hat{K}^{\dagger}|\ua\ran = - \lan\da|\hat{K}|\ua\ran\ ,
\end{equation}
which implies that for $\lan\ua|\hat{K}|\da\ran = 1$, $\lan\da|\hat{K}|\ua\ran = -1$. Using these we get the elemental inner products as
\bea
&& \lan\ua^A|\da^B\ran = \lan\da|\sigma_c^A\sigma_b^B|\ua\ran =
i\,\Omega^{BA}\lan\da|\hat{K}|\ua\ran = i\,\Omega^{AB}\ , \nonumber\\
&& \lan\da^A|\ua^B\ran = \lan\ua|\sigma_b^A\sigma_c^B|\da\ran =
i\,\Omega^{AB}\lan\ua|\hat{K}|\da\ran = i\,\Omega^{AB}\ .
\eea
It is then possible to construct ghost-spin chains with nearest neighbour
interactions between operators at neighbouring lattice sites, somewhat
similar to the ghost-spin chain for the $bc$-ghost CFTs. However the
continuum limit is less clear in this case, in part due to technical
difficulties such as the construction of the Jordan-Wigner
transformation to obtain fermionic versions of the $\sigma_{b,c}$
operators above which anticommute with each other (the $\sigma_{b,c}$
are bosonic spin-like operators commuting at neighbouring lattice sites
while anticommuting at the same site).
Note however that the case with $N=1$ has structure similar to that
appearing in the theory of anticommuting scalars: this is a
logarithmic CFT in 2-dimensions \cite{Gurarie:1993xq,Kausch:1995py,
  Gurarie:1997dw,Kausch:2000fu,Flohr:2001zs,Krohn:2002gh,Gurarie:2013tma}.
So perhaps the continuum limit here gives symplectic fermions\ $\int
\Omega_{AB}\ \del\phi^A\del\phi^B$: we hope to explore this further.

\section{$N$ irreducible levels}\label{sec:n-irreducible-levels}

In this section, we consider a generalization of ghost-spins that
consists of $N$ irreducible levels, defined as
\begin{equation}\label{irredN}
  \langle e_i|e_i\rangle=0\ ,  \qquad \langle e_i|e_j\rangle=1 \qquad
  \forall\quad i\neq j\ ; \qquad i,j=1,2,\dots,N\ .
\end{equation}
For $N=2$, the basis states $|e_1\ran, |e_2\ran$ are identical to the
$|\ua\ran, |\da\ran$ basis states, and this system reduces to the
2-level ghost-spin reviewed in Sec.~2.
Flavoured generalizations can be constructed by adding additional
flavour indices to these, along the lines we have described for 2-level
ghost-spins in the previous sections: we will not do so here however.

Using the inner products above, it is clear that there are various
negative norm states here as well: \eg\ $|e_i\ran-|e_j\ran$ has norm $-2$.
Using a diagonal basis helps as in the 2-level case to identify positive
and negative norm states clearly. This can be done using the
transformations in Appendix~\ref{Nirred-basistransform}: we can choose
an orthonormal basis where the basis states and their inner products are\
\begin{eqnarray}\label{irredNbasisSt}
&& |\alpha\ran\equiv \{|+\ran,\ |2\ran,\ \dots,\ |N\ran\}\ ;\nonumber\\
&&  \lan\alpha|\beta\ran = \eta_{\alpha\beta}\ ; \qquad \eta_{++}=1\ ,\quad \eta_{22} = \eta_{33} = \cdots \cdots = \eta_{NN}=-1\ , \quad \eta_{\alpha\beta} = 0\quad \forall\ \alpha\neq\beta\ ,\quad \nonumber \\
  \ie && \langle +|+\rangle=1\ , \qquad \langle \alpha|\alpha\rangle=-1 ,
  \qquad \alpha=2,...,N\ .
\end{eqnarray}
Then the generic state and its norm in both bases are
\begin{eqnarray}
&& |\psi\ran = \psi^i|e_i\ran; \qquad \lan\psi|\psi\ran = (\psi^i)^*\psi^j\lan e_i|e_j\ran = \sum_{i\neq j} (\psi^i)^*\psi^j\ , \\
&& |\psi\ran = \psi^{\alpha}|\alpha\ran; \qquad \lan\psi|\psi\ran = (\psi^{\alpha})^*\psi^{\beta}\lan \alpha|\beta\ran = (\psi^{\alpha})^*\psi^{\beta} \eta_{\alpha\beta} = |\psi^{+}|^2 - \sum_{\alpha=2}^N |\psi^{\alpha}|^2\ .\quad
\end{eqnarray}
To illustrate this, let us consider $N=3$. The generic state and its norm are
\begin{eqnarray}
&& |\psi\ran = \psi^1|e_1\ran + \psi^2|e_2\ran + \psi^3|e_3\ran = \psi^+|+\ran + \psi^2|2\ran + \psi^3|3\ran\ , \nonumber \\
&& \lan\psi|\psi\ran = (\psi^1)^*\psi^2 + (\psi^2)^*\psi^1 +(\psi^1)^*\psi^3 + (\psi^3)^*\psi^1 + (\psi^2)^*\psi^3 + (\psi^3)^*\psi^2 \nonumber\\
&& \qquad\quad = |\psi^{+}|^2 - |\psi^2|^2 - |\psi^3|^2\ .
\end{eqnarray}
In some sense, this is a ghost-spin generalization of the $N$-level
spins that arise in the Heisenberg spin chain: perhaps appropriate
interaction Hamiltonians for ghost-spin chains on a 1-dim lattice can
be studied along those lines.


\medskip

\noindent {\bf Correlated ghost-spins and entanglement:}\ \ 
We want to construct correlated ghost-spin states analogous to the
discussion in sec.~\ref{sec:CorrStO(N)}. So consider a system of
two ghost-spins with $N$ irreducible levels. The orthonormal basis
for this system is
\begin{equation}
  |u_A u_B\rangle\equiv |\alpha\ran|\beta\rangle \equiv |\al\beta\ran
  \qquad \forall\quad \alpha,\beta = +,2,\dots,N\ ,
\end{equation}
where each $|\al\ran$ is a single ghost-spin basis state in
(\ref{irredNbasisSt}).
A generic state $|\psi\rangle=\psi^{\alpha\beta}|\alpha\beta\rangle$ has a norm $\lan\psi|\psi\ran = \eta_{\alpha\kappa}\eta_{\beta\lambda}(\psi^{\alpha\beta})^*\psi^{\kappa\lambda}$, which can be expanded as
\begin{equation}
\langle\psi|\psi\rangle=\left(\sum_{\alpha}|\psi^{\alpha\alpha}|^2+\sum_{\alpha,\beta\neq +}|\psi^{\alpha\beta}|^2\right)-\left(\sum_{\alpha\neq +}(|\psi^{+\alpha}|^2+\psi^{\alpha+}|^2)\right)\ .
\end{equation}
We see a manifest division between the positive and negative norm
subspaces. For $N=3$, we can see this explicitly as
\begin{equation} \langle\psi|\psi\rangle=(|\psi^{++}|^2+|\psi^{22}|^2+|\psi^{33}|^2+|\psi^{23}|^2+|\psi^{32}|^2)-(|\psi^{+2}|^2+|\psi^{2+}|^2+|\psi^{+3}|^2+|\psi^{3+}|^2)\ .
\end{equation}
For general $N$, by tracing over the second ghost-spin, the reduced density matrix is
\begin{equation}
	\rho_A=(\rho_A)^{\alpha\kappa}|\alpha\rangle\langle\kappa|\ ; \qquad (\rho_A)^{\alpha\kappa} = \psi^{\alpha\beta}(\psi^{\kappa\beta})^*\eta_{\beta\beta}\ .
\end{equation}
The mixed index reduced density matrix is\
$(\rho_A)^{\alpha}_{\beta} = \eta_{\beta\kappa}(\rho_A)^{\kappa\alpha} = \eta_{\beta\kappa}\eta_{\lambda\lambda}\psi^{\kappa\lambda}(\psi^{\alpha\lambda})^*$.

\bigskip

\noindent\underline{Correlated ghost-spins\,:}\ \
From the norm above we see that the states $|++\ran$, $|22\ran$, $\dots$, $|NN\ran$ span the subspace of correlated ghost-spin states, where a generic correlated ghost-spin state is
\begin{equation}
	|\psi\ran=\psi^{\alpha\alpha}|\alpha\alpha\ran\ ;\qquad \lan\psi|\psi\ran = |\psi^{++}|^2 + |\psi^{22}|^2 + \cdots + |\psi^{NN}|^2\ .
\end{equation}

\bigskip

\noindent\underline{Entanglement pattern in a general state\,:}\ \
Consider a slightly more general state
\begin{equation}
|\psi\rangle=\sum_{\alpha=+}^{N}\psi^{\alpha\alpha}|\alpha\alpha\rangle+\sum_{\beta=2}^{N}(\psi^{+\beta}|+\beta\rangle+\psi^{\beta+}|\beta+\rangle)\ ,
\end{equation}
whose norm is
\begin{equation}
\langle\psi|\psi\rangle=\sum_{\alpha=+}^{N}|\psi^{\alpha\alpha}|^2-\sum_{\beta=2}^{N}(|\psi^{+\beta}|^2+|\psi^{\beta+}|^2)\ .
\end{equation}
The off-diagonal components of the reduced density matrix are
\begin{align}
(\rho_A)^{+\alpha} &=\psi^{++}\psi^{\alpha+^*}-\psi^{+\alpha}\psi^{\alpha\alpha^*}\ , \qquad \forall\ \alpha=2,\dots,N\ ,\ \alpha\neq +\ , \nonumber \\
(\rho_A)^{\alpha\beta} &=\psi^{\alpha+}\psi^{\beta+^*}\ , \qquad\qquad\qquad \forall\ \alpha,\beta=2,\dots,N\ ,\ \alpha\neq+\ ,\ \beta\neq+\ .
\end{align}
From $(\rho_A)^{\alpha\beta}=0$, we see that only one of $\psi^{\alpha+}$ is non-zero, i.e., $\psi^{2+}\neq 0$, $\psi^{\alpha+}=0$, $\alpha=3,\dots,N$. Then $(\rho_A)^{+\alpha}=0$ gives $\psi^{+2}\neq 0$ and $\psi^{+\alpha}=0$, $\alpha=3,\dots,N$.

So we consider the state
\begin{equation}
|\psi\rangle=\psi^{++}|++\rangle+\psi^{22}|22\rangle+\cdots+\psi^{NN}|NN\rangle+\psi^{+2}|+2\rangle+\psi^{2+}|2+\rangle\ ,
\end{equation}
whose norm is
\begin{equation}
\langle\psi|\psi\rangle=|\psi^{++}|^2+\,\cdots\,+|\psi^{NN}|^2-|\psi^{+2}|^2-|\psi^{2+}|^2\ .
\end{equation}
The non-zero components of the reduced density matrix are
\begin{eqnarray}
	&& (\rho_A)^{++} =|\psi^{++}|^2-|\psi^{+2}|^2\ , \qquad (\rho_A)^{22} =|\psi^{2+}|^2-|\psi^{22}|^2\ , \nonumber \\
	&& (\rho_A)^{+2} =\psi^{++}\psi^{2+^*}-\psi^{+2}\psi^{22^*}\ , \qquad (\rho_A)^{\alpha\alpha} =-|\psi^{\alpha\alpha}|^2\ , \quad \alpha=3,\dots,N\ .
\end{eqnarray}
Choosing $\psi^{2+^*}=\frac{\psi^{+2}\psi^{22^*}}{\psi^{++}}$ and defining $x\equiv |\psi^{++}|^2-|\psi^{+2}|^2$ and $r\equiv \dfrac{|\psi^{22}|^2}{|\psi^{++}|^2}>0$, the mixed-index components of $\rho_A$ are
\begin{equation}
(\rho_A)^+_+=x\ , \quad (\rho_A)^2_2=xr\ , \quad (\rho_A)^{\alpha}_{\alpha}=|\psi^{\alpha\alpha}|^2\ ,\quad \alpha=3,\dots,N
\end{equation}
and
\begin{equation}
\langle\psi|\psi\rangle=x+xr+|\psi^{33}|^2+\cdots+|\psi^{NN}|^2=\pm 1\ .
\end{equation}
Now depending on if $x$ is positive or negative we have the following three cases.

\noindent $\bullet $ If $x>0$, $|\psi\ran$ has necessarily positive norm and $\langle\psi|\psi\rangle=1$ implies $0<(\rho_A)^{\alpha}_{\alpha}<1$ for all $\alpha=+,2,\dots,N$ giving $S_A>0$.

\medskip

\noindent $\bullet$ If $x<0$, the norm of $|\psi\ran$ can be positive or negative.

\noindent (i) For positive norm, \ie\ $\langle\psi|\psi\rangle=-|x|-|x|r+\sum_{\alpha=3}^{N}|\psi^{\alpha\alpha}|^2=1$, we get
\begin{equation}\begin{split}
S_A&=|x|\log|x|+|x|r\log|x|r-|\psi^{33}|^2\log(|\psi^{33}|^2)-\sum_{\alpha=4}^{N}|\psi^{\alpha\alpha}|^2\log(|\psi^{\alpha\alpha}|^2)+i\pi|x|(1+r) \\
&=|x|\log|x|+|x|r\log|x|r-\left(1+|x|+|x|r-\sum_{\alpha=4}^{N}|\psi^{\alpha\alpha}|^2\right)\log\left(1+|x|+|x|r-\sum_{\alpha=4}^{N}|\psi^{\alpha\alpha}|^2\right) \\
&\qquad -\sum_{\alpha=4}^{N}|\psi^{\alpha\alpha}|^2\log(|\psi^{\alpha\alpha}|^2)+i\pi|x|(1+r)\ .
\end{split}\end{equation}
We see that $Im(S_A)$ is not constant and $Re(S_A)<0$ when
\begin{eqnarray}
|x|\log|x|+|x|r\log|x|r &<& \Big(1+|x|+|x|r-\sum_{\alpha=4}^{N}|\psi^{\alpha\alpha}|^2\Big)\log\Big(1+|x|+|x|r+\sum_{\alpha=4}^{N}|\psi^{\alpha\alpha}|^2\Big) \nonumber \\
&& \hspace{4cm} +\sum_{\alpha=4}^{N}|\psi^{\alpha\alpha}|^2\log(|\psi^{\alpha\alpha}|^2)\ .
\end{eqnarray}

\noindent (ii) For a negative norm state $|\psi\ran$, \ie\ $\langle\psi|\psi\rangle=-|x|-|x|r+\sum_{\alpha=3}^{N}|\psi^{\alpha\alpha}|^2=-1$, we get
\begin{eqnarray}
	S_A &=& |x|\log|x|+|x|r\log|x|r-|\psi^{33}|^2\log(|\psi^{33}|^2)-\sum_{\alpha=4}^{N}|\psi^{\alpha\alpha}|^2\log(|\psi^{\alpha\alpha}|^2)+i\pi|x|(1+r) \nonumber \\
	&=& |x|\log|x|+|x|r\log|x|r -\sum_{\alpha=4}^{N}|\psi^{\alpha\alpha}|^2\log(|\psi^{\alpha\alpha}|^2)+i\pi|x|(1+r) \nonumber\\
	&&\quad -\left(-1+|x|+|x|r-\sum_{\alpha=4}^{N}|\psi^{\alpha\alpha}|^2\right)\log\left(-1+|x|+|x|r-\sum_{\alpha=4}^{N}|\psi^{\alpha\alpha}|^2\right)
\end{eqnarray}
$Im(S_A)$ is constant if $|x|+|x|r=c$, where $c$ is a constant and $c>1$ (from the norm). Then $Re(S_A)$ becomes
\begin{eqnarray}
	Re(S_A) &=& |x|\log|x|+(c-|x|)\log(c-|x|)-\Big(c-1-\sum_{\alpha=4}^{N}|\psi^{\alpha\alpha}|^2\Big)\log\Big(c-1-\sum_{\alpha=4}^{N}|\psi^{\alpha\alpha}|^2\Big) \nonumber\\
	&&\qquad -\sum_{\alpha=4}^{N}|\psi^{\alpha\alpha}|^2\log(|\psi^{\alpha\alpha}|^2)\ .
\end{eqnarray}
We see that $Re(S_A)<0$ for those values of $|x|$, $c>1$, $\psi^{\alpha\alpha}$ which satisfy
\begin{equation}
|x|\log|x|+(c-|x|)\log(c-|x|)<\Big(c-1-\sum_{\alpha=4}^{N}|\psi^{\alpha\alpha}|^2\Big)\log\Big(c-1-\sum_{\alpha=4}^{N}|\psi^{\alpha\alpha}|^2\Big)+\sum_{\alpha=4}^{N}|\psi^{\alpha\alpha}|^2\log(|\psi^{\alpha\alpha}|^2)\ .
\end{equation}

\section{Discussion}\label{sec:discussion}

We have constructed $N$-level generalizations of the 2-level
ghost-spins in \cite{Narayan:2016xwq,Jatkar:2016lzq,Jatkar:2017jwz}.
These include (i) a flavoured generalization comprising $N$ copies of
the ghost-spin system and corresponding ghost-spin chains which lead
to 2-dim $bc$-ghost CFTs with $O(N)$) flavour symmetry, (ii) a
spin-glass type coupling in flavour space, (iii) a symplectic
generalization involving antisymmetric inner products between the
elemental ghost-spins, and (iv) an irreducible ghost-spin system with
$N$ internal levels. We have studied entanglement properties in these
cases: among other things, these show the existence of positive norm
states in two copies of ghost-spin ensembles obtained by entangling
identical ghost-spins from each copy: these are akin to the correlated
ghost-spin states in \cite{Jatkar:2016lzq,Jatkar:2017jwz}, and exhibit
positive entanglement.

We now describe briefly some of the motivations from $dS/CFT$, in
particular \cite{Narayan:2017xca}, for the studies here.
Generalizations of gauge/gravity duality for de Sitter space or
$dS/CFT$ involve conjectured dual hypothetical Euclidean non-unitary
CFTs living on the future boundary ${\cal I}^+$
\cite{Strominger:2001pn,Witten:2001kn, Maldacena:2002vr}. Using the
dictionary $\Psi_{dS}=Z_{CFT}$\ \cite{Maldacena:2002vr}, where
$\Psi_{dS}$ is the late-time Hartle-Hawking wavefunction of the
universe with appropriate boundary conditions and $Z_{CFT}$ the dual
CFT partition function, the dual CFT$_d$ energy-momentum tensor
correlators reveal central charge coefficients ${\cal C}_d\sim
i^{1-d}{l^{d-1}\over G_{d+1}}$ in $dS_{d+1}$ (effectively analytic
continuations from $AdS/CFT$). This is real and negative in $dS_4$,
with \,${\cal C}_3\sim -{R_{dS}^2\over G_4}$\, so that $dS_4/CFT_3$
is reminiscent of ghost-like non-unitary theories.
Bulk expectation values are of the form\
$\langle \varphi_k \varphi_{k'}\rangle  \sim \int D\varphi\ \varphi_k
\varphi_{k'} |\Psi_{dS}|^2$. This involves the probability
$|\Psi_{dS}|^2=\Psi_{dS}^*\Psi_{dS}$, which suggests that bulk de Sitter
physics involves two copies of the dual CFT --- $CFT_F\times CFT_P$ on
the future and past boundaries. This is unlike in
$AdS/CFT$ where $Z_{bulk}=Z_{bndry}$ implies boundary correlators 
can be obtained as a limit of bulk ones. In the $dS$ case, while
$Z_{CFT}=\Psi_{dS}$ of a single dual CFT copy at $I^+$ can be
used to obtain boundary correlators (\eg\ $\lan O_kO_{k'}\ran\sim
{\delta^2Z_{{}_{CFT}}\over\delta\varphi_k^0\delta\varphi_{k'}^0}$\
for operators $O_k$ dual to modes $\varphi_k$), bulk observables require
$|\Psi_{dS}|^2$, the bulk probability, and so two copies of the dual
$Z_{CFT}$. This dovetails with the structure of extremal
surfaces and entanglement as we see below.

In $AdS$, surfaces anchored at one end of a subsystem dip into the
bulk radial direction and then begin to return to the boundary at
turning points. In $dS$, the boundary at $I^+$ is spacelike and
surfaces dip into the time direction which ends up making their
structure quite different, as studied in \cite{Narayan:2015vda}. For
instance, considering the $dS$ Poincare slicing
$ds^2={R_{dS}^2\over\tau^2}(-d\tau^2+dx_i^2)$\,, a strip subsystem on
some boundary Euclidean time $w=const$ slice of $I^+$ with width along
$x$ gives a bulk extremal surface $x(\tau)$ described by\ ${\dot
  x}^2\equiv ({dx\over d\tau})^2 = {B^2\tau^{2d-2}\over
  1+B^2\tau^{2d-2}}$\ ($B^2>0$). $w$, $x$ can be taken as any of the $x_i$
(so the boundary Euclidean time slice is not sacrosanct).
Compared with the $AdS$ case, the denominator here crucially has a
relative minus sign.  Thus there is no \emph{real} turning point here
where the surface starting at $I^+$ begins to turn back towards
$I^+$:\ this requires $|{\dot x}|\ra\infty$ while here $|{\dot x}|\leq
1$. There are also complex extremal surfaces however, which exhibit turning
points: these end up amounting to analytic continuation from the $AdS$
Ryu-Takayanagi surfaces.  While their interpretation is not entirely
conclusive, in $dS_4$ they have \emph{negative} area, consistent with
the negative central charge in $dS_4/CFT_3$ as mentioned above.

Since surfaces starting at $I^+$ do not turn back, it is then
interesting to ask if they could instead stretch all the way to the
past boundary $I^-$.
In \cite{Narayan:2017xca}, connected codim-2 extremal surfaces in the
static patch coordinatization of de Sitter space were found stretching
from $I^+$ to $I^-$ passing through the vicinity of the bifurcation
region with divergent area ${l^2\over 4G_4} {1\over\epsilon}$, where
$\epsilon={\epsilon_c\over l}$ is the dimensionless ultraviolet cutoff
and the coefficient scales as de Sitter entropy.
To elaborate a little, the static patch coordinatization can be
recast as\
$ds^2 = {l^2\over\tau^2} \big(-{d\tau^2\over 1-\tau^2} + (1-\tau^2) dw^2
+ d\Omega_{d-1}^2\big)$, with the future/past universes $F/P$
parametrized by $0\leq \tau\leq 1$ with horizons at $\tau=1$, while
the Northern/Southern diamonds $N/S$ have $1<\tau\leq\infty$. The
boundaries at $\tau=0$ are now of the form $R_w\times S^{d-1}$,
resembling the Poincare slicing locally.
Setting up the extremization for codim-2 surfaces on boundary Euclidean
time slices can be carried out: on $S^{d-1}$ equatorial planes for
instance we obtain\
${\dot w}^2 = {B^2\tau^{2d-2}\over 1-\tau^2 + B^2\tau^{2d-2}}$\,. The
minus sign here, reflecting the horizons, makes the structure of these
surfaces interesting, drawing parallels with the $AdS$ extremization
(we refer to \cite{Narayan:2017xca} for further details).
The limit $B\ra 0$ gives surfaces passing through the vicinity of the
bifurcation region as stated above, with the width $\Delta w$ approaching
all of $I^\pm$. These connected surfaces stretching between $I^\pm$ are
akin to rotated versions of the connected surfaces of Hartman, Maldacena
\cite{Hartman:2013qma} in the AdS black hole. This led to the
speculation there that $dS_4$ is approximately dual to an entangled
thermofield-double type state of the form
\be\label{dsEntstate}
|\psi\rangle = \sum \psi^{i_n^F,i_n^P} |i_n^F\ran |i_n^P\ran
\ee
akin to the thermofield double \cite{Maldacena:2001kr} dual to $AdS$
black holes.
Here $\psi^{i_n^F,i_n^P}$ are coefficients entangling a generic ghost-spin
$|i_n^F\ran$ from $CFT_F$ at $I^+$ with an identical one $|i_n^P\ran$
from $CFT_P$ at $I^-$. The constituent states are schematically
continuum versions of $N$ level ghost-spins, with $N$ related to
$dS_4$ entropy ${l^2\over 4G_4}$\,. Since bulk time evolution maps
configurations at $I^-$ to those at $I^+$ \cite{Witten:2001kn}, we have
the schematic map\ $|i_n^P\ran \ra S[i_n^P, i_n^F] |i_n^P\ran \equiv
|i_n^F\ran$ where $S[i_n^P, i_n^F]$ is the operator representing
bulk time evolution\ (note that this is a bulk object that is to be
distinguished from operators in the $CFT$ representing boundary Euclidean
time evolution). If states $|i^P\ran$ at $I^-$ map faithfully and completely
to states $|i^P\ran$ at $I^+$, then $S$ is expected to be a unitary
operator (this is also vindicated by the fact that exchanging $I^\pm$
is a bulk symmetry).
This suggests that the entangled states (\ref{dsEntstate}) are
unitarily equivalent to similar maximally entangled states
$|\psi\rangle = \sum \psi^{i_n^F,i_n^F} |i_n^F\ran |i_n^F\ran$
in two $CFT_F$ copies of the ghost-CFT solely at $I^+$.\
The state (\ref{dsEntstate}) is akin to a correlated ghost-spin state
with an even number of ghost-spins, as discussed in
\cite{Jatkar:2016lzq,Jatkar:2017jwz}.  It necessarily has positive
norm\
$\sum_{i_n^F,i_n^F} \gamma_{i_n^F,i_n^F}\ \gamma_{i_n^F,i_n^F}\ \psi^{i_n^F,i_n^F}
(\psi^{i_n^F,i_n^F})^*  \ra \ \sum_{i_n} |\psi^{i_n^F,i_n^F}|^2$,\
since we are entangling identical
states $i_n^F$ and $i_n^P$: thus it has positive entanglement, as in
\cite{Jatkar:2016lzq,Jatkar:2017jwz}. Since each constituent state
$|i_n^{F,P}\ran$ is $N$-level, \ie\ with $N$ internal degrees of
freedom, the entanglement entropy scales as\ $N\sim {l^2\over G_4}$\,.
The toy models in sec.~\ref{sec:CorrStO(N)} of correlated ghost-spin
states (\ref{corrStgsc}) and their entanglement entropy (\ref{corrStEE})
are of this form, written explicitly.
The state (\ref{dsEntstate}) is akin to the thermofield double dual to
the eternal $AdS$ black hole \cite{Maldacena:2001kr}. This suggests
the speculation that 4-dim de Sitter space is perhaps approximately
dual to $CFT_F\times CFT_P$ in the entangled state (\ref{dsEntstate})
and the generalized entanglement entropy of the latter scales as de
Sitter entropy. (See \cite{Dong:2018cuv} for another approach
to de Sitter entropy based on the $dS/dS$ correspondence
\cite{Alishahiha:2004md}.)

The investigations in this paper on $N$-level generalizations of
ghost-spins are geared towards constructing microscopic
ghost-spin states that reflect the $N$-level internal structure which
might ultimately give rise in appropriate continuum limits to
theories such as the $Sp(N)$ ghost-CFT dual to higher spin $dS_4$
(see also the recent work \cite{Anninos:2017eib}).
As we have seen, the $N$-level generalizations here do admit positive
norm subsectors of the form of the correlated ghost-spin states
indicated in (\ref{dsEntstate}).

As mentioned in the Introduction, the ghost-spin system has possible
applications in gauge theories.  The continuum limit of a $d$-dim
ghost-spin system (as in \cite{Jatkar:2017jwz} for the 2-dim case)
with flavour quantum numbers may be relevant for studying entanglement
in gauge theories in a covariant setting.  A better understanding of
the ghost-spin system and its coupling to ordinary spin systems as in
\cite{Jatkar:2016lzq,Jatkar:2017jwz} generalized to $d$-dimensions
would be an ideal sandbox for understanding covariant formulations of
subregion entanglement in gauge theories.

We have been thinking of ghost-spins as microscopic building blocks
for ghost-like CFTs, and perhaps more general non-unitary CFTs. The
discussions in this paper on ghost-spin chains have recovered 2-dim
$bc$-ghost CFTs with flavour symmetries.  The obvious generalization
to 3 dimensions of the 2-dim case discussed in sec.3 (and in
\cite{Jatkar:2017jwz} for the single flavour case) has nearest
neighbour hopping-type interactions of the elemental
form\ $h\sim\sum_{nn'}\sigma_{b,{\vec n}}^A\sigma_{c,{\vec n'}}^A$\,.
This contains three $\sigma_b$ and three $\sigma_c$ operators at each
lattice site, with possible flavour indices reflecting internal
flavour symmetries.  It would be interesting to study such 3-dim
ghost-spin chains towards obtaining 3-dim ghost-CFTs in the continuum
limit: so far, we have encountered conceptual difficulties as well as
technical ones. We hope to report on this in the future.

\vspace{5mm}

{\footnotesize \noindent {\bf Acknowledgements:}\ \ It is a pleasure
  to thank Philip Argyres, Jan de Boer, R. Loganayagam, Juan
  Maldacena, Rob Myers, Eva Silverstein and Tadashi Takayanagi for
  useful conversations on \cite{Narayan:2017xca} and some of the
  content in this paper.  While this work was in progress, KK thanks
  the organizers of the PiTP school ``From Qubits to spacetime'', IAS,
  Princeton, Jul 2018; KN thanks the organizers of ``Strings 2018'',
  Okinawa, and the longterm Yukawa memorial workshop ``New Frontiers
  in String Theory'', Yukawa Institute, Kyoto, Japan, Jun-Jul
  2018. The research of KK and KN was supported in part by the
  International Centre for Theoretical Sciences (ICTS) during a visit
  for participating in the program - AdS/CFT at 20 and Beyond (Code:
  ICTS/adscft20/05).  We thank the organizers of the Indian Strings
  Meeting ISM2018, IISER Trivandrum, for hospitality as this work was
  being finalized. This work is partially supported by a grant to CMI
  from the Infosys Foundation.  }

\appendix

\section{Orthonormal basis for $N$-level irreducible ghost-spin}\label{Nirred-basistransform}

We give the transformations which transform the defining basis for $N$-level irreducible ghost-spin to an orthonormal basis. For even $N$-level ghost-spin, the transformations to orthonormal basis are
\begin{eqnarray}
  && |+\rangle=\frac{\sum_{i=1}^{N}|e_i\rangle}{\sqrt{N^2-N}}\ , \quad |2\rangle=\frac{1}{\sqrt{2}}(|e_1\rangle-|e_2\rangle)\ , \quad \dots,\quad \left|\frac{N}{2}+1\right\rangle=\frac{1}{\sqrt{2}}(|e_{N-1}\rangle-|e_{N}\rangle)\ , \nonumber\\
  && \left|\frac{N}{2}+2\right\rangle=\frac{1}{2}(|e_1\rangle+|e_2\rangle-|e_3\rangle-|e_4\rangle)\ , \quad  \left|\frac{N}{2}+3\right\rangle=\frac{1}{\sqrt{12}}\Big(\sum_{i=1}^{4}|e_i\rangle-2(|e_5\rangle+|e_6\rangle)\Big)\ , \nonumber \\
  && \vdots \nonumber \\
  && |N-1\rangle =\frac{1}{\sqrt{\frac{(N-1)(N-3)}{2}}}\left(\sum_{i=1}^{N-4}|e_i\rangle-\left(\frac{N-1}{2}-1\right)(|e_{N-3}\rangle+|e_{N-2}\rangle)\right)\ , \nonumber \\
  && |N\rangle =\frac{1}{\sqrt{\frac{N(N-2)}{2}}}\left(\sum_{i=1}^{N-2}|e_i\rangle-\left(\frac{N}{2}-1\right)(|e_{N-1}\rangle+|e_{N}\rangle)\right)\ .
\end{eqnarray}
For odd $N+1$-level ghost-spin (where $N$ is even), the transformations to orthonormal basis as
\begin{eqnarray}
  && |+\rangle=\frac{\sum_{i=1}^{N+1}|e_i\rangle}{\sqrt{N(N+1)}}\ , \quad |2\rangle=\frac{1}{\sqrt{2}}(|e_1\rangle-|e_2\rangle)\ , \quad \dots, \quad \left|\frac{N}{2}+1\right\rangle=\frac{1}{\sqrt{2}}(|e_{N-1}\rangle-|e_{N}\rangle)\ ,\nonumber \\
  && \left|\frac{N}{2}+2\right\rangle=\frac{1}{2}(|e_1\rangle+|e_2\rangle-|e_3\rangle-|e_4\rangle)\ , \quad \left|\frac{N}{2}+3\right\rangle=\frac{1}{\sqrt{12}}\Big(\sum_{i=1}^{4}|e_i\rangle-2(|e_5\rangle+|e_6\rangle)\Big)\ ,\nonumber \\
  && \vdots \nonumber \\
  && |N-1\rangle=\frac{1}{\sqrt{\frac{(N-1)(N-3)}{2}}}\left(\sum_{i=1}^{N-4}|e_i\rangle-\left(\frac{N-1}{2}-1\right)(|e_{N-3}\rangle+|e_{N-2}\rangle)\right)\ , \nonumber \\
  && |N\rangle=\frac{1}{\sqrt{\frac{N(N-2)}{2}}}\left(\sum_{i=1}^{N-2}|e_i\rangle-\left(\frac{N}{2}-1\right)(|e_{N-1}\rangle+|e_{N}\rangle)\right)\ , \nonumber \\
  && |N+1\rangle=\frac{1}{\sqrt{(N+1)^2-(N+1)}}\left(\sum_{i=1}^{N}|e_i\rangle-N|e_{N+1}\rangle\right)\ .
\end{eqnarray}
To illustrate these transformation, we write them explicitly for $N=3$ and $N=4$. For a $3$-level ghost-spin, the orthonormal basis are
\begin{equation}
  |+\rangle=\frac{1}{\sqrt{6}}(|e_1\rangle+|e_2\rangle+|e_3\rangle)\ , \quad |2\rangle=\frac{1}{\sqrt{2}}(|e_1\rangle-|e_2\rangle)\ , \quad |3\rangle =\frac{1}{\sqrt{6}}(|e_1\rangle+|e_2\rangle-2|e_3\rangle)
\end{equation}
and for $N=4$ the orthonormal basis are
\begin{eqnarray}
  && |+\rangle=\frac{1}{\sqrt{12}}(|e_1\rangle+|e_2\rangle+|e_3\rangle+|e_1\rangle)\ , \qquad |2\rangle=\frac{1}{\sqrt{2}}(|e_1\rangle-|e_2\rangle)\ , \nonumber \\
  && |3\rangle=\frac{1}{\sqrt{2}}(|e_3\rangle-|e_4\rangle)\ , \qquad |4\rangle=\frac{1}{\sqrt{2}}(|e_1\rangle+|e_2\rangle-|e_3\rangle-|e_4\rangle)\ .
\end{eqnarray}


\begin{thebibliography}{}
  
{ 
  
\footnotesize{

 \bibitem{Narayan:2016xwq}
 K.~Narayan,
 ``On $dS_4$ extremal surfaces and entanglement entropy in some ghost CFTs,''
 Phys.\ Rev.\ D {\bf 94}, no. 4, 046001 (2016)
 doi:10.1103/PhysRevD.94.046001
 [arXiv:1602.06505 [hep-th]].

 \bibitem{Jatkar:2016lzq}
 D.~P.~Jatkar and K.~Narayan,
 ``Entangled spins and ghost-spins,''
 Nucl.\ Phys.\ B {\bf 922}, 319 (2017)
 doi:10.1016/j.nuclphysb.2017.07.002
 [arXiv:1608.08351 [hep-th]].
 
 \bibitem{Jatkar:2017jwz}
 D.~P.~Jatkar and K.~Narayan,
 ``Ghost-spin chains, entanglement and $bc$-ghost CFTs,''
 Phys.\ Rev.\ D {\bf 96}, no. 10, 106015 (2017)
 doi:10.1103/PhysRevD.96.106015
 [arXiv:1706.06828 [hep-th]].

\bibitem{Strominger:2001pn} 
  A.~Strominger,
  ``The dS / CFT correspondence,''
  JHEP {\bf 0110}, 034 (2001)
  [hep-th/0106113].

\bibitem{Witten:2001kn} 
  E.~Witten,
  ``Quantum gravity in de Sitter space,''
  [hep-th/0106109].

\bibitem{Maldacena:2002vr}
  J.~M.~Maldacena,
  ``Non-Gaussian features of primordial fluctuations in single field inflationary models,''
  JHEP {\bf 0305}, 013 (2003),\ 
  [astro-ph/0210603].

\bibitem{Anninos:2011ui} 
  D.~Anninos, T.~Hartman and A.~Strominger,
  ``Higher Spin Realization of the dS/CFT Correspondence,''
  arXiv:1108.5735 [hep-th].

\bibitem{Bousso:2001mw} 
  R.~Bousso, A.~Maloney and A.~Strominger,
  ``Conformal vacua and entropy in de Sitter space,''
  Phys.\ Rev.\ D {\bf 65}, 104039 (2002)
  doi:10.1103/PhysRevD.65.104039
  [hep-th/0112218].
  
\bibitem{Balasubramanian:2002zh} 
  V.~Balasubramanian, J.~de Boer and D.~Minic,
  ``Notes on de Sitter space and holography,''
  Class.\ Quant.\ Grav.\  {\bf 19}, 5655 (2002)
  [Annals Phys.\  {\bf 303}, 59 (2003)]
  [hep-th/0207245].

\bibitem{Harlow:2011ke} 
  D.~Harlow and D.~Stanford,
  ``Operator Dictionaries and Wave Functions in AdS/CFT and dS/CFT,''
  arXiv:1104.2621 [hep-th].

\bibitem{Ng:2012xp} 
  G.~S.~Ng and A.~Strominger,
  ``State/Operator Correspondence in Higher-Spin dS/CFT,''
  Class.\ Quant.\ Grav.\  {\bf 30}, 104002 (2013)
  [arXiv:1204.1057 [hep-th]].

\bibitem{Das:2012dt} 
  D.~Das, S.~R.~Das, A.~Jevicki and Q.~Ye,
  ``Bi-local Construction of Sp(2N)/dS Higher Spin Correspondence,''
  JHEP {\bf 1301}, 107 (2013)
  [arXiv:1205.5776 [hep-th]].

\bibitem{Anninos:2012ft} 
  D.~Anninos, F.~Denef and D.~Harlow,
  ``The Wave Function of Vasiliev's Universe - A Few Slices Thereof,''
  Phys.\ Rev.\ D {\bf 88}, 084049 (2013)
  [arXiv:1207.5517 [hep-th]].

\bibitem{Anninos:2017eib} 
  D.~Anninos, F.~Denef, R.~Monten and Z.~Sun,
  ``Higher Spin de Sitter Hilbert Space,''
  arXiv:1711.10037 [hep-th].

 \bibitem{Narayan:2017xca} 
 K.~Narayan,
 ``On extremal surfaces and de Sitter entropy,''
 Phys.\ Lett.\ B {\bf 779}, 214 (2018)
 doi:10.1016/j.physletb.2018.02.010
 [arXiv:1711.01107 [hep-th]].

\bibitem{Narayan:2015vda} 
  K.~Narayan,
  ``de Sitter extremal surfaces,''
  Phys.\ Rev.\ D {\bf 91}, no. 12, 126011 (2015)
  doi:10.1103/PhysRevD.91.126011
  [arXiv:1501.03019 [hep-th]].

\bibitem{Narayan:2015oka} 
  K.~Narayan,
  ``de Sitter space and extremal surfaces for spheres,''
  Phys.\ Lett.\ B {\bf 753}, 308 (2016)
  doi:10.1016/j.physletb.2015.12.019
  [arXiv:1504.07430 [hep-th]].

\bibitem{Ryu:2006bv} 
  S.~Ryu and T.~Takayanagi,
  ``Holographic derivation of entanglement entropy from AdS/CFT,''
  Phys.\ Rev.\ Lett.\  {\bf 96}, 181602 (2006)
  [hep-th/0603001].

\bibitem{Ryu:2006ef} 
  S.~Ryu and T.~Takayanagi,
  ``Aspects of Holographic Entanglement Entropy,''
  JHEP {\bf 0608}, 045 (2006)
  [hep-th/0605073].

\bibitem{HRT} 
V.~E.~Hubeny, M.~Rangamani and T.~Takayanagi,
``A Covariant holographic entanglement entropy proposal,'' 
JHEP {\bf 0707} (2007) 062  [arXiv:0705.0016 [hep-th]].

\bibitem{Rangamani:2016dms} 
  M.~Rangamani and T.~Takayanagi,
  ``Holographic Entanglement Entropy,''
  Lect.\ Notes Phys.\  {\bf 931}, pp.1 (2017)
  doi:10.1007/978-3-319-52573-0
  [arXiv:1609.01287 [hep-th]].

\bibitem{LeClair:2006kb} 
  A.~LeClair,
  ``Quantum critical spin liquids, the 3D Ising model, and conformal field theory in 2+1 dimensions,''
  cond-mat/0610639.

\bibitem{LeClair:2007iy} 
  A.~LeClair and M.~Neubert,
  ``Semi-Lorentz invariance, unitarity, and critical exponents of symplectic fermion models,''
  JHEP {\bf 0710}, 027 (2007)
  doi:10.1088/1126-6708/2007/10/027
  [arXiv:0705.4657 [hep-th]].

\bibitem{Gurarie:1993xq} 
  V.~Gurarie,
  ``Logarithmic operators in conformal field theory,''
  Nucl.\ Phys.\ B {\bf 410}, 535 (1993)
  doi:10.1016/0550-3213(93)90528-W
  [hep-th/9303160].

\bibitem{Kausch:1995py} 
  H.~G.~Kausch,
  ``Curiosities at c = -2,''
  hep-th/9510149.

\bibitem{Gurarie:1997dw} 
  V.~Gurarie, M.~Flohr and C.~Nayak,
  ``The Haldane-Rezayi quantum Hall state and conformal field theory,''
  Nucl.\ Phys.\ B {\bf 498}, 513 (1997)
  doi:10.1016/S0550-3213(97)00351-9
  [cond-mat/9701212].

\bibitem{Kausch:2000fu} 
  H.~G.~Kausch,
  ``Symplectic fermions,''
  Nucl.\ Phys.\ B {\bf 583}, 513 (2000)
  [hep-th/0003029].

\bibitem{Flohr:2001zs} 
  M.~Flohr,
  ``Bits and pieces in logarithmic conformal field theory,''
  Int.\ J.\ Mod.\ Phys.\ A {\bf 18}, 4497 (2003)
  doi:10.1142/S0217751X03016859
  [hep-th/0111228].

\bibitem{Krohn:2002gh} 
  M.~Krohn and M.~Flohr,
  ``Ghost systems revisited: Modified Virasoro generators and logarithmic conformal field theories,''
  JHEP {\bf 0301}, 020 (2003)
  doi:10.1088/1126-6708/2003/01/020
  [hep-th/0212016].

\bibitem{Gurarie:2013tma} 
  V.~Gurarie,
  ``Logarithmic operators and logarithmic conformal field theories,''
  J.\ Phys.\ A {\bf 46}, 494003 (2013)
  doi:10.1088/1751-8113/46/49/494003
  [arXiv:1303.1113 [cond-mat.stat-mech]].

\bibitem{Hartman:2013qma} 
  T.~Hartman and J.~Maldacena,
  ``Time Evolution of Entanglement Entropy from Black Hole Interiors,''
  JHEP {\bf 1305}, 014 (2013)
  doi:10.1007/JHEP05(2013)014
  [arXiv:1303.1080 [hep-th]].

\bibitem{Maldacena:2001kr} 
  J.~M.~Maldacena,
  ``Eternal black holes in anti-de Sitter,''
  JHEP {\bf 0304}, 021 (2003)
  doi:10.1088/1126-6708/2003/04/021
  [hep-th/0106112].

\bibitem{Dong:2018cuv} 
  X.~Dong, E.~Silverstein and G.~Torroba,
  ``De Sitter Holography and Entanglement Entropy,''
  JHEP {\bf 1807}, 050 (2018)
  doi:10.1007/JHEP07(2018)050
  [arXiv:1804.08623 [hep-th]].

\bibitem{Alishahiha:2004md} 
  M.~Alishahiha, A.~Karch, E.~Silverstein and D.~Tong,
  ``The dS/dS correspondence,''
  AIP Conf.\ Proc.\  {\bf 743}, 393 (2005)
  doi:10.1063/1.1848341
  [hep-th/0407125].



}
}
\end{thebibliography}
\end{document}